\documentclass[12pt]{article}
\pdfoutput=1
\usepackage{amssymb, amsmath,amsfonts}
\usepackage{multirow}
\usepackage{mathrsfs}
\usepackage{array}
\usepackage{cite}
\usepackage{booktabs}
\usepackage{tikz}
\usepackage{pgfplots}
\usepackage{float}
\usepackage{subfigure, graphicx}
\usepackage[pdftex, bookmarks=true,colorlinks,linkcolor=red,urlcolor=blue,citecolor=blue]{hyperref}
\usepackage{cite}
\usepackage{slashed}
\usepackage{tabularx}

\usepackage{tikz}

\makeatletter\@addtoreset{equation}{section}\makeatother

\def\be{\begin{equation}}
\def\ee{\end{equation}}
\def\bea{\begin{eqnarray}}
\def\eea{\end{eqnarray}}

\newcommand{\nn}{\nonumber}

\def\Dslash{\,\,{\raise.15ex\hbox{/}\mkern-12mu D}}
\def\Dbarslash{\,\,{\raise.15ex\hbox{/}\mkern-12mu {\bar D}}}
\def\delslash{\,\,{\raise.15ex\hbox{/}\mkern-9mu \partial}}
\def\delbarslash{\,\,{\raise.15ex\hbox{/}\mkern-9mu {\bar\partial}}}
\def\pslash{\,\,{\raise.15ex\hbox{/}\mkern-9mu p}}
\def\calDslash{\,\,{\raise.15ex\hbox{/}\mkern-12mu {\cal D}}}

\makeatletter\@addtoreset{equation}{section}\makeatother

\renewcommand{\title}[1]{\vbox{\center\LARGE{#1}}\vspace{5mm}}
\renewcommand{\author}[1]{\vbox{\center#1}\vspace{5mm}}
\newcommand{\address}[1]{\vbox{\center\em#1}}

\def\arXiv#1{\href{http://arxiv.org/abs/#1}{arXiv:#1}}
\def\arXiv#1#2{\href{http://arxiv.org/abs/#1}{arXiv:#1}}

\def\doi#1#2{\href{http://doi.org/#1}{#2}}

\begin{document}

\unitlength = .8mm

\begin{titlepage}
\vspace{.5cm}
 
\begin{center}
\hfill \\
\hfill \\
\vskip 1cm

\title{\boldmath 
Black hole interiors \\in holographic topological semimetals
}
\vskip 0.5cm
{Ling-Long Gao$^{\,a,b}$}\footnote{Email: {\tt linglonggao@buaa.edu.cn}},
{Yan Liu$^{\,a,b}$}\footnote{Email: {\tt yanliu@buaa.edu.cn}} and
{Hong-Da Lyu$^{\,a,b}$}\footnote{Email: {\tt hongdalyu@buaa.edu.cn}}

\address{${}^{a}$Center for Gravitational Physics, Department of Space Science\\ and International Research Institute
of Multidisciplinary Science, \\Beihang University, Beijing 100191, China}

\address{${}^{b}$Peng Huanwu Collaborative Center for Research and Education, \\Beihang University, Beijing 100191, China}

\end{center}
\vskip 1.5cm

\abstract{We study the black hole interiors in 
holographic Weyl semimetals and holographic nodal line semimetals. We find that the black hole singularities are of Kasner form. In the topologically nontrivial phase at low temperature, both the Kasner exponents of the metric fields and the proper time from the horizon to the singularity are almost constant, likely reflecting the topological nature of the topological semimetals. We also find some specific behaviors inside the horizon in  each holographic semimetal model.}
\vfill

\end{titlepage}

\begingroup 
\hypersetup{linkcolor=black}
\tableofcontents
\endgroup




\section{Introduction}

The conventional classifications on the phases of matter are rooted in the Landau paradigm of symmetry breaking theory \cite{McGreevy:2022oyu}. Over the past thirty years, new states of matter have been found which are beyond the concept of Landau paradigm. One example is the topological states of matter, including the quantum Hall states, topological insulators, topological semimetals and so on \cite{wen}. 
Different from the conventional Landau paradigm, there is no symmetry breaking during the topological phase transition and it attracts lots of research attention.

In recent years, the strongly interacting topological Weyl semimetals (WSM) \cite{Landsteiner:2015lsa, Landsteiner:2015pdh} and nodal line semimetals (NLSM) \cite{Liu:2018bye, Liu:2020ymx} have been explicitly constructed from the holographic duality. Both holographic WSM and NLSM are shown to possess nontrivial topological invariants \cite{Liu:2018djq}. Remarkably, the holographic WSM exhibits interesting effects inherited from the boundary states \cite{Ammon:2016mwa}. These features indicate that the physical properties associated to topology in the weakly coupled field theories 
persist in the strongly coupled topological systems from the holography. Moreover, the systems could go through a topological phase transition to a topologically trivial semimetal phase, see \cite{Landsteiner:2019kxb} for a review on the developments.\footnote{Other interesting developments can be found in e.g. \cite{Landsteiner:2016stv, Copetti:2016ewq, Grignani:2016wyz, Ammon:2018wzb, Baggioli:2018afg, Liu:2018spp, Ji:2019pxx, Song:2019asj, Juricic:2020sgg, Baggioli:2020cld, Kiczek:2020qsw, BitaghsirFadafan:2020lkh, Zhao:2021pih, Grandi:2021bsp, Rodgers:2021azg, Ji:2021aan, Zhao:2021qfo, Grandi:2021jkj}. } In the holographic WSM, during the topological phase transition the anomalous Hall conductivity can serve as an order parameter, while in the holographic NLSM it is not clear which observable could serve as an order parameter.  Whether possible universal ``order parameter" exist for the topological phase transitions? What is the topological nature in the topological phase from holography? These are elusive problems  we aim to explore from the holographic duality.  

In holography, the thermal states are dual to black hole geometries in the bulk. The black hole interior is expected to encode important information of the dual field theory \cite{Fidkowski:2003nf, Festuccia:2005pi, Grinberg:2020fdj}. In the case that the thermal states are described by black holes with simple Kasner singularities, it has been shown recently in \cite{Liu:2021hap} that the order of
the thermal phase transition in the dual field theory is connected to the behavior of the Kasner exponents of the black hole singularity.\footnote{Other studies on the geometric aspects of  black hole singularities can be found in e.g. \cite{Frenkel:2020ysx,  Hartnoll:2020rwq, Hartnoll:2020fhc, Liu:2022rsy, 
Cai:2020wrp, Wang:2020nkd, An:2021plu, Grandi:2021ajl, Mansoori:2021wxf, Dias:2021afz, Sword:2021pfm, Cai:2021obq, Henneaux:2022ijt, Caceres:2022smh, Bhattacharya:2021nqj, An:2022lvo, 
Auzzi:2022bfd, Mirjalali:2022wrg, Hartnoll:2022snh, Caceres:2022hei, Hartnoll:2022rdv, Sword:2022oyg}. } For the topological phase transitions in holographic topological semimetals at finite temperature, the systems experience a smooth crossover from a topological phase, a critical phase to a trivial phase. Although the phase crossover is different from thermal phase transitions,   it is still interesting to 
explore 
the interior geometries in holographic topological semimetals, in order to uncover possible universal behavior during the topological phase transitions.

It turns out that there exist both universal and  special behaviors of the singularities in holographic topological semimetals. The universal behavior is similar to the topological nature of the topological phase and might give hints to the problems we raised for topological semimetals, while the special features can be understood from the fact that the holographic WSM and the holographic NLSM share similarities and also differences in the constructions as emphasized in \cite{Liu:2018bye, Liu:2018djq}. More precisely, in both cases, two matter fields are added which play the same role from the point of view of the boundary field theory, while they play different roles in the bulk geometry. In the boundary field theory, one of the two matter fields is to deform the Dirac point into two Weyl nodes or a nodal line, while the other matter field is to gap the system. In the bulk, in the topological phase of the holographic WSM the IR geometry of Schwarzschild black hole is not deformed by the matter fields, while in the topological phase of the holographic NLSM  the backreaction of the matter fields on the gravitational geometry is quite strong in IR. We will see that these two different situations lead to different properties of the black hole singularities in the topological phases.

It is known that the information of the interior geometry can be probed from the geodesics which correspond to certain correlators in the dual field theory. For example, the proper time from the horizon to the singularity can be extracted from the thermal one point function of certain heavy operator \cite{Grinberg:2020fdj}. We will compute this quantity in the bulk and study its behavior in the topological phases and trivial phases.

This paper is organized as follows. In Sec. \ref{sec2}, we will first review the holographic WSM and then study its interior geometry as well as the proper time of the timelike geodesics. In Sec. \ref{sec3}, we will  review the holographic NLSM and then also study its interior geometry and the proper time of the timelike geodesics. Sec. \ref{sec4} is devoted to the conclusions and open questions. The details of calculations are in the appendices.

\section{Inside holographic Weyl semimetal}
\label{sec2}

In this section we first briefly review the holographic WSM which describes a topological phase transition from topological WSM phase to a trivial semimetal phase. Then we study the interior geometry of the black hole solutions and discuss the possible universal behavior of the black hole singularities as well as the interior geometry. We also comment on the possible observable as the role of ``order parameter" during the topological phase transition. 

The action of the holographic WSM \cite{Landsteiner:2015lsa, Landsteiner:2015pdh} is
\begin{align}
\label{eq:actionwsm}
\begin{split}
  S=& \int d^5x \sqrt{-g}\, \bigg[ \frac{1}{2\kappa^2}\big(R+\frac{12}{L^2}\big) -\frac{1}{4}\mathcal{F}^2 -\frac{1}{4}F^2 +\frac{\alpha}{3} \epsilon^{abcde} A_{a} \big( F_{bc}F_{de} +3 \mathcal{F}_{bc} \mathcal{F}_{de} \big) \\
   &~~~~~- (D_{a}\Phi)^*(D^{a}\Phi)-V(\Phi)  \bigg]\,,
   \end{split}
\end{align}
where two gauge fields are dual to vector and axial currents respectively. A special Chern-Simons structure is introduced to match the Wald identity for these currents. An axially charged scalar field $\Phi$ is also introduced in the model with the source interpreted as the mass term. Note that $D_a\Phi=\partial_a\Phi - iq A_a\Phi$ where $A_a$ is the axial $U(1)$ gauge potential, and $V(\Phi)=m^2|\Phi|^2+\frac{\lambda}{2}|\Phi|^4$.  We set $2\kappa^2=L=1$.

We focus on the finite temperature and use the following ansatz
\begin{align}
\label{eq:ansatz}
\begin{split}
  ds^2 &= -udt^2+\frac{dr^2}{u}+ f(dx^2+dy^2)+hdz^2\,,\\
  A &= A_z dz\,, \quad \Phi=\phi\,.
\end{split}
\end{align}
The equations of motion for the fields can be found in appendix \ref{app:hwsm}.
In the following we consider $m^2=-3, q=1, \lambda=1/10.$ Generalization to other values of the parameters is straightforward. 

We use the following boundary conditions for the matter fields
\be
\lim_{r\to\infty}A_z=b\,,~~~\lim_{r\to\infty} r\phi=M\,,
\ee
where $b$ is the time reversal symmetry breaking parameter which plays the role of splitting a Dirac point into two Weyl points, and $M$ is the mass parameter which gaps the Dirac point. The competition between these two effects leads to interesting topological phase transitions. The system is completely determined by the dimensionless parameters $T/b, M/b$. 

\begin{figure}[h!]
\begin{center}
\includegraphics[width=0.57\textwidth]{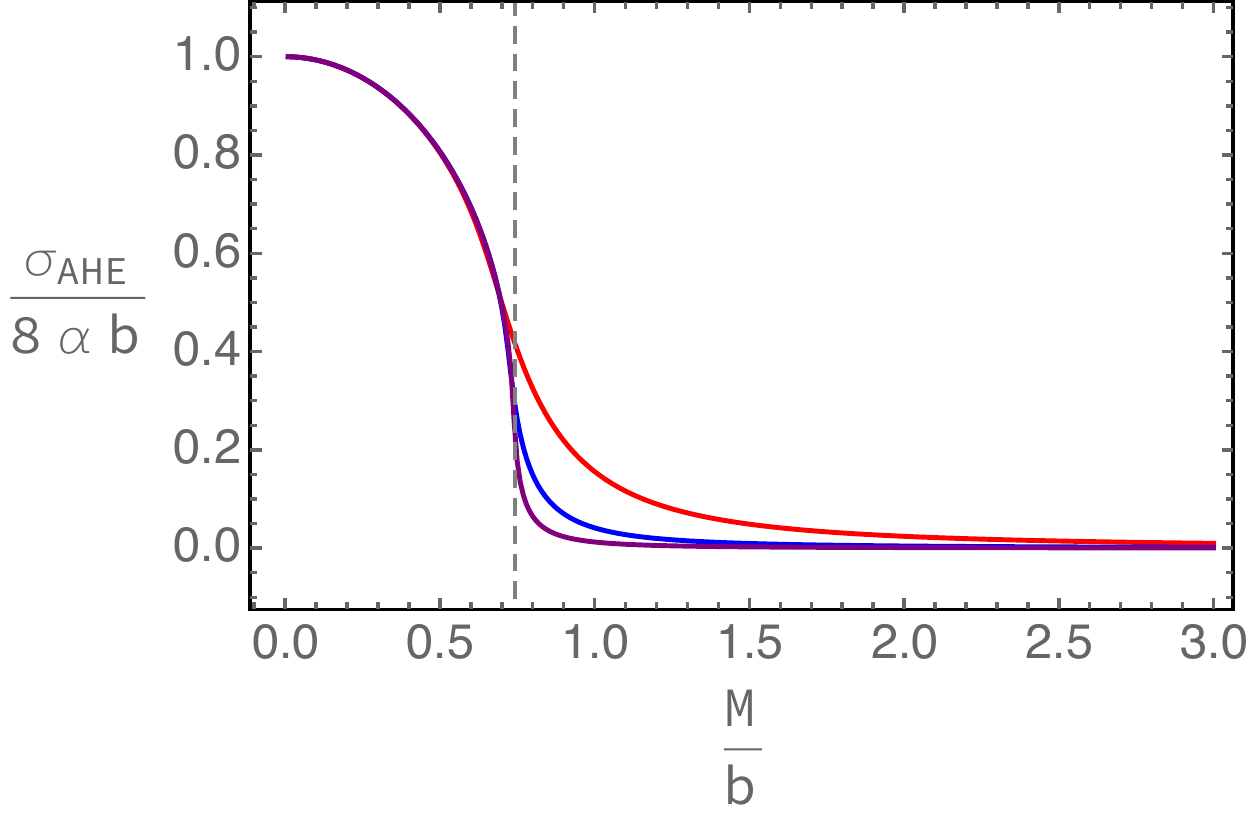}
~~
\end{center}
\vspace{-0.3cm}
\caption{\small  
Plot of anomalous Hall conductivity as a function of $M/b$  
at the temperatures $T/b=0.05$ (red), $0.02$ (blue), $0.01$ (purple). The gray dashed line is the critical value of $M/b$ of the quantum phase transition at zero temperature.}
\label{fig:ahe}
\end{figure}

In the weakly coupled WSM, the quantum topological phase transition could be manifested  from the band structure and equivalently the behavior of the anomalous Hall conductivity. In the strongly coupled model from holography, the anomalous Hall conductivity behaves similarly to the weakly coupled case, indicating that there is a topological phase transition, as shown in Fig. \ref{fig:ahe}. The lines in red, blue and purple are for $T/b=0.05, 0.02, 0.01$ respectively. The transition becomes sharp at zero temperature and the dashed gray line is the critical value of the transition $(M/b)_c\simeq 0.744$  at zero temperature.

\subsection{Inner structures}
\label{sec:ist}
The phase transitions can be parameterized by the anomalous Hall conductivity which is completely determined by the horizon value of the axial gauge field $A_z$. Given the possible connection between the physics inside and outside the horizon, it is interesting to study the black hole inner structures during the topological phase transitions. 

\begin{figure}[h!]
\begin{center}
\includegraphics[width=0.46\textwidth]{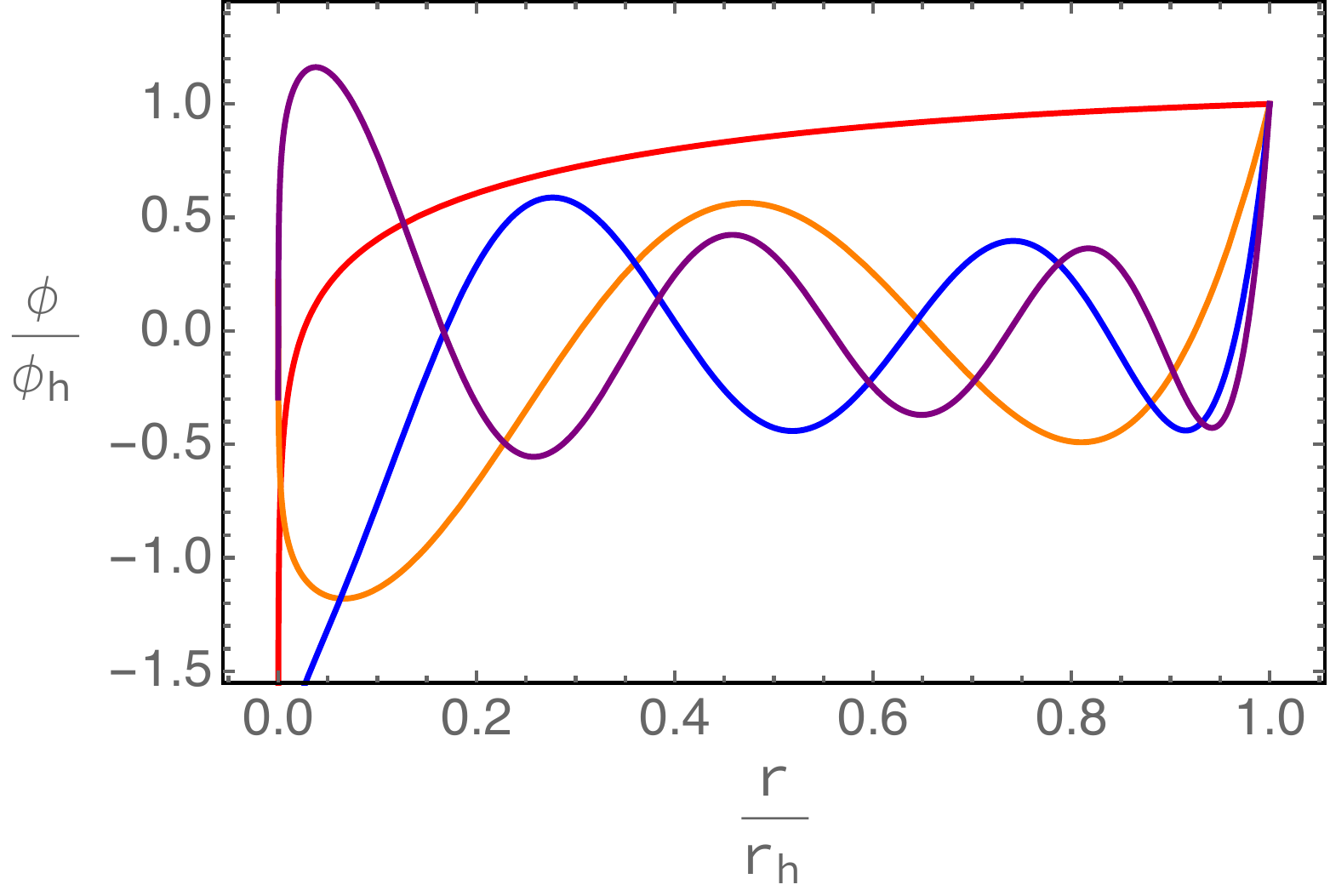}
~~
\includegraphics[width=0.46\textwidth]{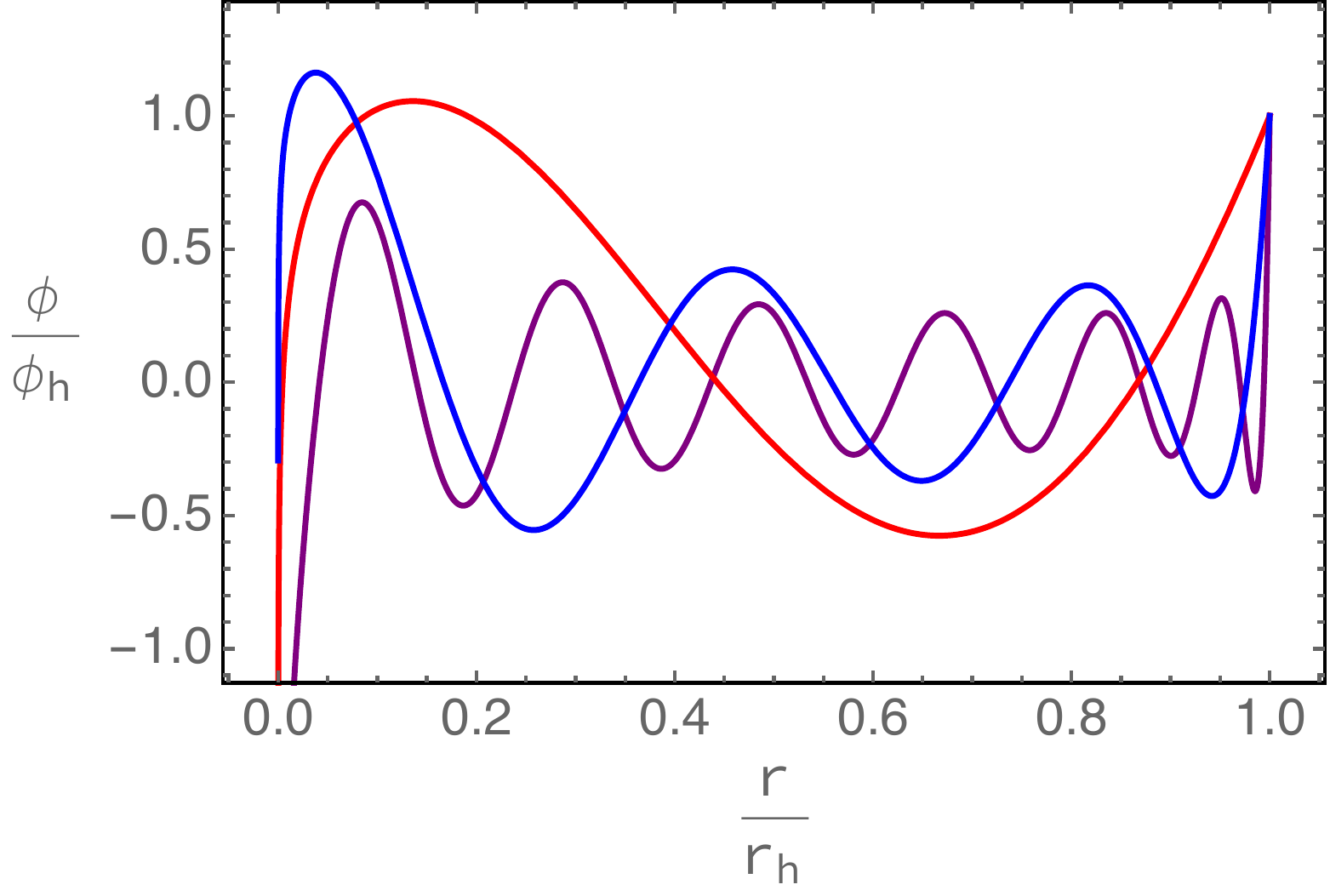}
\end{center}
\vspace{-0.3cm}
\caption{\small  The plots of $\phi/\phi_h$ along radial direction in the oscillation region at  fixed $T/b=0.02$ (left) while $M/b = 0.1$ (purple), $0.4$ (blue), $0.6$ (orange), $0.74$ (red), as well as at fixed $M/b=0.1$ (right) while $T/b=0.05$ (red), $0.02$ (blue), $0.01$ (purple). Here $\phi_h$ is the horizon value of $\phi$.
}
\label{fig:phiosc}
\end{figure}

From the black hole solution we have obtained, we can  integrate the system further to the singularity since the geometry is smooth at the horizon. We find that at low temperature, the matter field $\phi$ oscillates along the  direction $r$ inside the horizon only in the topological phase (i.e. $M/b<0.744$). The typical behavior 
is shown in Fig. \ref{fig:phiosc}, where the profiles of the scalar field $\phi$ (which have been rescaled according to $\phi/\phi_h$) in the oscillation regime as a function of $r/r_h$ at fixed  $T/b$ (left) or $M/b$ (right) are plotted respectively. We find that when we fix the temperature $T/b$, the times of oscillations become less when we increase $M/b$ from $0$ to $(M/b)_c$. When we fix $M/b< (M/b)_c$, $\phi$ oscillates more times at lower  temperature. Note that the other fields do not show any oscillation from the horizon to the singularity. 

Different from the holographic superconductor cases, the oscillation here is not related to the collapse of Einstein-Rosen bridge \cite{Hartnoll:2020fhc}, since there is no inner horizon any more for holographic WSM.\footnote{Similar oscillation behavior has been found previously in neutral helical black holes \cite{Liu:2022rsy}.} Note that in the holographic superconductor \cite{Hartnoll:2020fhc}, the oscillations of the scalar field are similar to the Josephson effect since the equation for the scalar field can be solved analytically under certain approximation, which means that the oscillation of the stiffness is determined by a background phase winding. However, in our case we have not found any similar physics. The approximate analytical solution of the scalar field is not related to a background phase winding. Nevertheless, we can still solve analytically the equation of motion for the scalar field near the singularity and the horizon 
and find that they match well with the numerical solution at low temperature and small $M/b$. The details can be found in appendix \ref{app:osc}.

\subsection{Behaviors of Kasner exponents}
\label{subsec:wsmkas}
The interior solution can be further integrated to the singularity. Near the singularity $r_s$, we assume that at the leading order the fields behave as 
\begin{align}
\label{eq:asns}
    u\sim -u_0 (r-r_s)^{n_u}\,,~~~ f\sim f_0 (r-r_s)^{n_f}\,,~~~ h\sim h_0 (r-r_s)^{n_h}\,,~~~ \phi\sim n_{\phi} \ln (r-r_s)\,,
\end{align}
where $u_0, f_0, h_0$ and $n_u, n_f, n_h, n_\phi$ are all constants. Here $u_0, f_0, h_0$ depend on the scaling symmetry in \eqref{eq:scasym1}, \eqref{eq:scasym2}, \eqref{eq:scasym3} while $n_u, n_f, n_h, n_\phi$ are not. Also note that here $r_s$ is not necessarily to be zero since there is a shift symmetry of the system $r\to r+\alpha$ along the radial direction which was used to set the boundary behavior \eqref{eq:nbwsm}. Moreover, as we shall see later, the axial gauge field $A_z$ is determined by the ansatz \eqref{eq:asns}.   

Near the singularity the equations of motion (\ref{eq:eom}) can be simplified under the assumption that the ignored terms are subleading which will be checked numerically  afterward, 
\begin{align}\label{eq:eomns}
\begin{split}
     u''+\frac{h'}{2h}u'-\left( f''+ \frac{f'h'}{2h}  \right)\frac{u}{f}&=0\,,\\
     \frac{f''}{f}+\frac{u''}{2u}-\frac{f'^2}{4f^2} + \frac{f'u'}{fu}+\frac{1}{2} \phi'^2 &=0\,,\\
    \frac{1}{2}\phi'^2 - \frac{u'}{2u}\left( \frac{f'}{f} + \frac{h'}{2h} \right)  - \frac{f'h'}{2fh} - \frac{f'^2}{4f^2}&=0\,,\\
    A_z''+\left( \frac{f'}{f} - \frac{h'}{2h}+ \frac{u'}{u} \right)A_z'&=0\,,\\
    \phi'' + \left( \frac{f'}{f}+\frac{h'}{2h}+ \frac{u'}{u} \right)\phi'&=0\,.
\end{split}
\end{align}
Substituting \eqref{eq:asns} into \eqref{eq:eomns}, we obtain
\begin{align}
\label{eq:relations}
n_h=2\,(1-n_u-n_f)\,, ~~~~~
n_{\phi}=\pm \sqrt{(2n_f+n_u)(1-n_u)- \frac{3n_f^2}{2}} \,.
\end{align}
We can also solve the fourth equation in  \eqref{eq:eomns} to obtain at leading order $A_z$, 
\begin{align}
\label{eq:nhazwsm}
    A_z \simeq A_{zs0} +  A_{zs1} (r-r_s)^{n_h} \,.
\end{align}
Note that the leading term $A_{zs0}$ can be rescaled to be $1$, while $A_{zs1}$ could be determined from the radially  conserved quantities as will be discussed later. 
Thus there are only two independent parameters in \eqref{eq:asns} and \eqref{eq:nhazwsm}. 

Note that in the above equations \eqref{eq:eomns}, we have assumed that the terms ignored are 
subleading. More explicitly, we have assumed
\begin{align}
\label{eq:paregime}
    n_u<2\,,~~~n_f+n_u<1\,, ~~~
    2 n_f + n_u>0\,.
\end{align}
Numerically we have checked that all the above relations are satisfied for the parameters we have considered, which indicates that the singularities are stable and of form \eqref{eq:asns} and \eqref{eq:nhazwsm}. 

There are two radially conserved charge associated to the scaling symmetries of the system, 
\bea
\label{eq:cc1}
    Q_1&=&\sqrt{h}(u'f-uf')\,,\\
\label{eq:cc2}
    Q_2&=& u'\sqrt{h}f- \frac{h'}{\sqrt{h}}uf -A_z A_z' \frac{uf}{\sqrt{h}}\,.
\eea
We have used them to check the accuracy of the numerics. Moreover, 
evaluating them at the horizon and at the singularity we obtain
\bea\label{eq:cc1s}
    4\pi T f_1 \sqrt{h_1} =Ts &=& u_0 f_0 \sqrt{h_0} (n_f-n_u)\\
&=&   \frac{u_0f_0}{\sqrt{h_0}} \left( n_h A_{zs0}A_{zs1} - h_0 (2n_f+3n_u-2) \,\right)
\eea
where $s$ is the entropy density. 
From \eqref{eq:cc1s}, we have $n_f>n_u$ in addition to the constraints \eqref{eq:paregime}. 
Moreover, the above two conserved quantities give the relations $n_h A_{zs0}A_{zs1}=h_0(3n_f+2n_u-2)$ which turns out to be zero in the topological phase at low temperature where $A_{zs1}=3n_f+2n_u-2=0$. 

Starting from (\ref{eq:ansatz}, \ref{eq:asns}) and performing a coordinate transformation  
\be\label{eq:coortr} 
\tau=-\frac{2}{\sqrt{n_0}(n_u-2)}(r-r_s)^{(2-n_u)/2}\,,\ee 
we obtain the Kasner form for the fields
\begin{align}
\label{eq:kasnerform}
\begin{split}
    ds^2 &= -d\tau^2 + c_t \tau^{2p_t} dt^2 + c_x \tau^{2p_x} (dx^2+dy^2) + c_z \tau^{2p_z} dz^2\,,\\
    \phi &=  p_{\phi}\log\tau + c_{\phi}\,,
\end{split}
\end{align}
where
\begin{align}
    \begin{split}
        p_t &= \frac{n_u}{2-n_u}\,,~~~~ p_x = \frac{n_f}{2-n_u}\,,~~~~
        p_z = \frac{n_h}{2-n_u}\,, ~~~~p_{\phi} =\frac{2 n_{\phi}}{2-n_u} \,.
    \end{split}
\end{align}
Note that $A_z$ is a constant at the leading order. 
Using the relations \eqref{eq:relations}, the above Kasner exponents can be expressed in terms of $n_u$ and $n_f$, 
\begin{align}
\label{eq:ptnurel}
\begin{split}
    &p_t = \frac{n_u}{2-n_u}\,,~~p_x = \frac{n_f}{2-n_u}\,,~~
    p_z =\frac{2(1-n_u-n_f)}{2-n_u}\,, ~~
    p_{\phi} = \pm \frac{\sqrt{4(2n_f+n_u)(1-n_u)-6n_f^2}}{2-n_u}\,.
\end{split}
\end{align}
Note that the sign of $p_\phi$ in \eqref{eq:ptnurel} can only be determined from numerics.  
They satisfy the following Kasner relations
\begin{align}
\label{eq:kasrel}
    p_t+2p_x+p_z=1\,,~~~~~
    p_t^2 + 2p_x^2+ p_z^2 +p_{\phi}^2=1\,.
\end{align}
It indicates that only two of the four Kasner exponents are independent. 

\begin{figure}[h!]
\begin{center}
\includegraphics[width=0.444\textwidth]{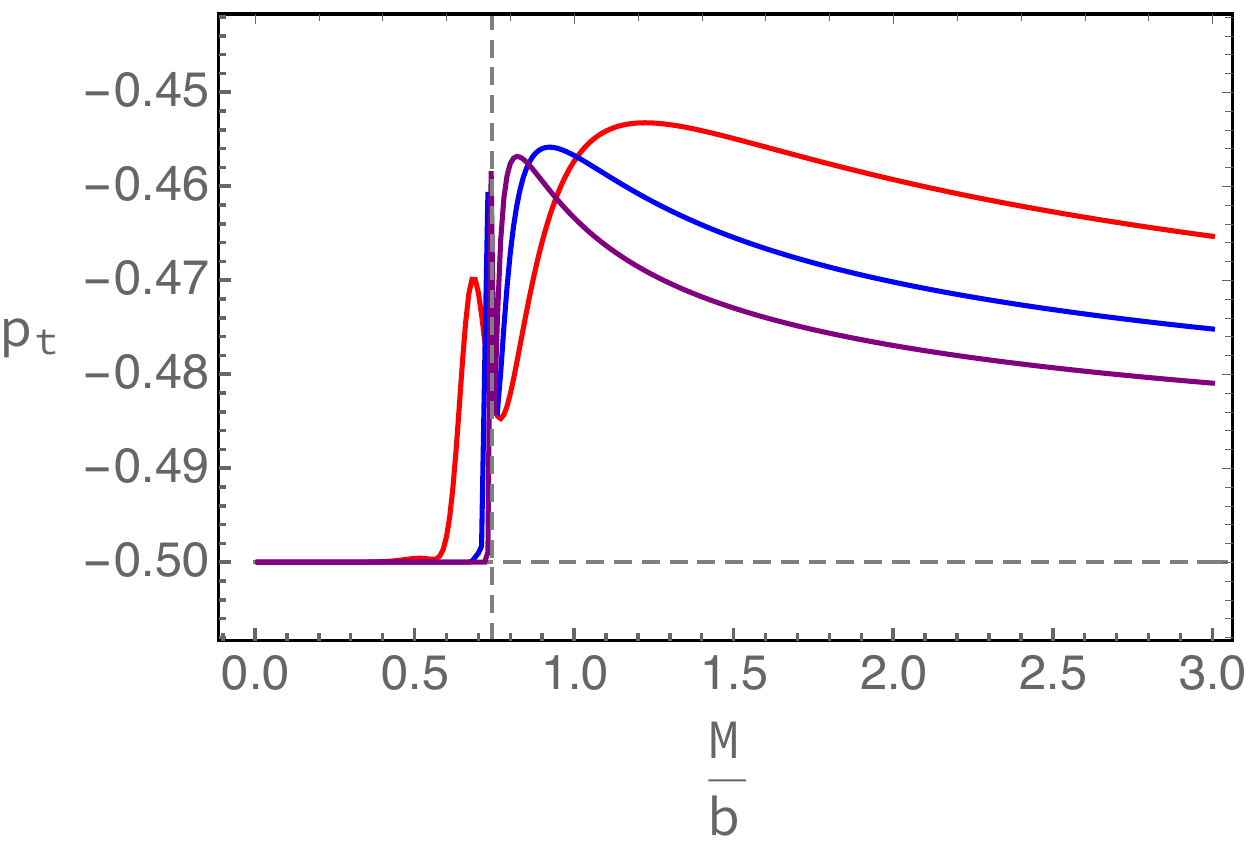}~~
\includegraphics[width=0.444\textwidth]{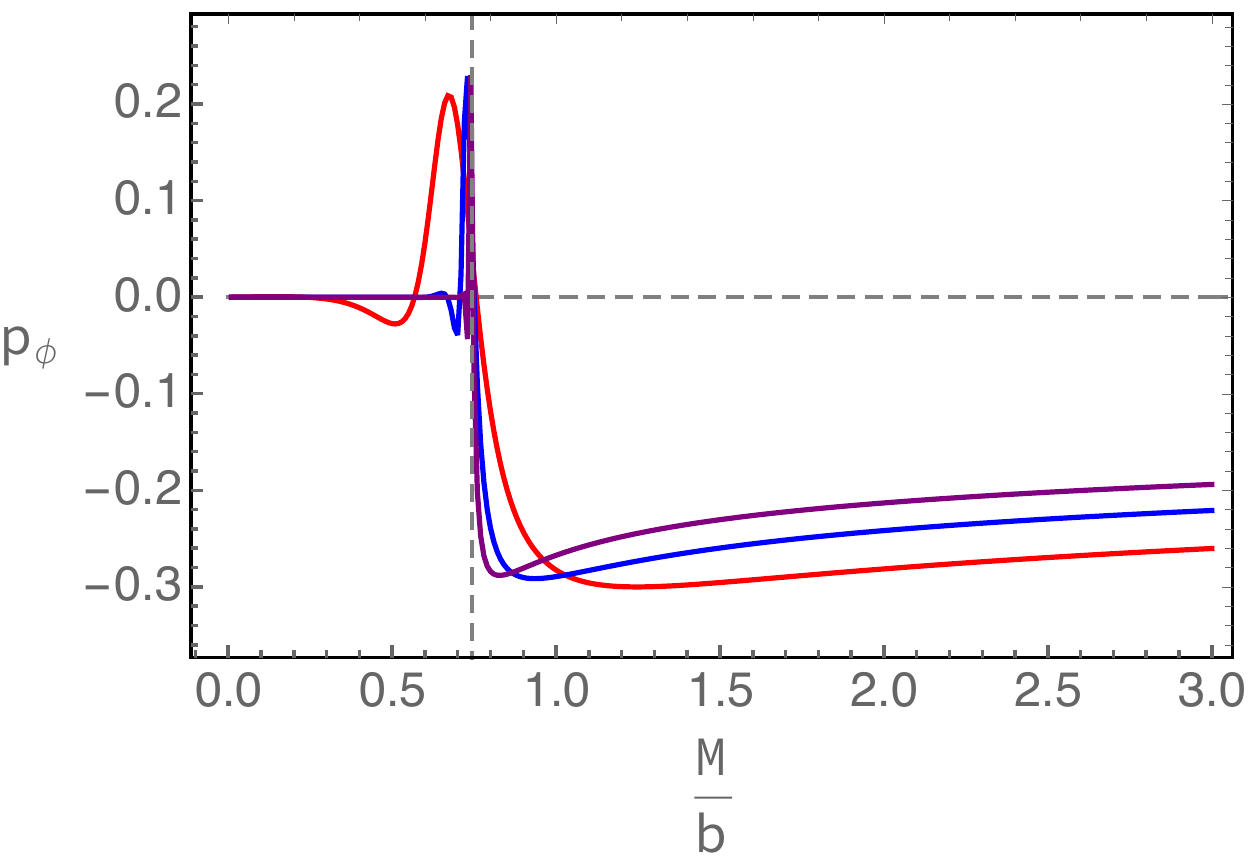}\\
\includegraphics[width=0.444\textwidth]{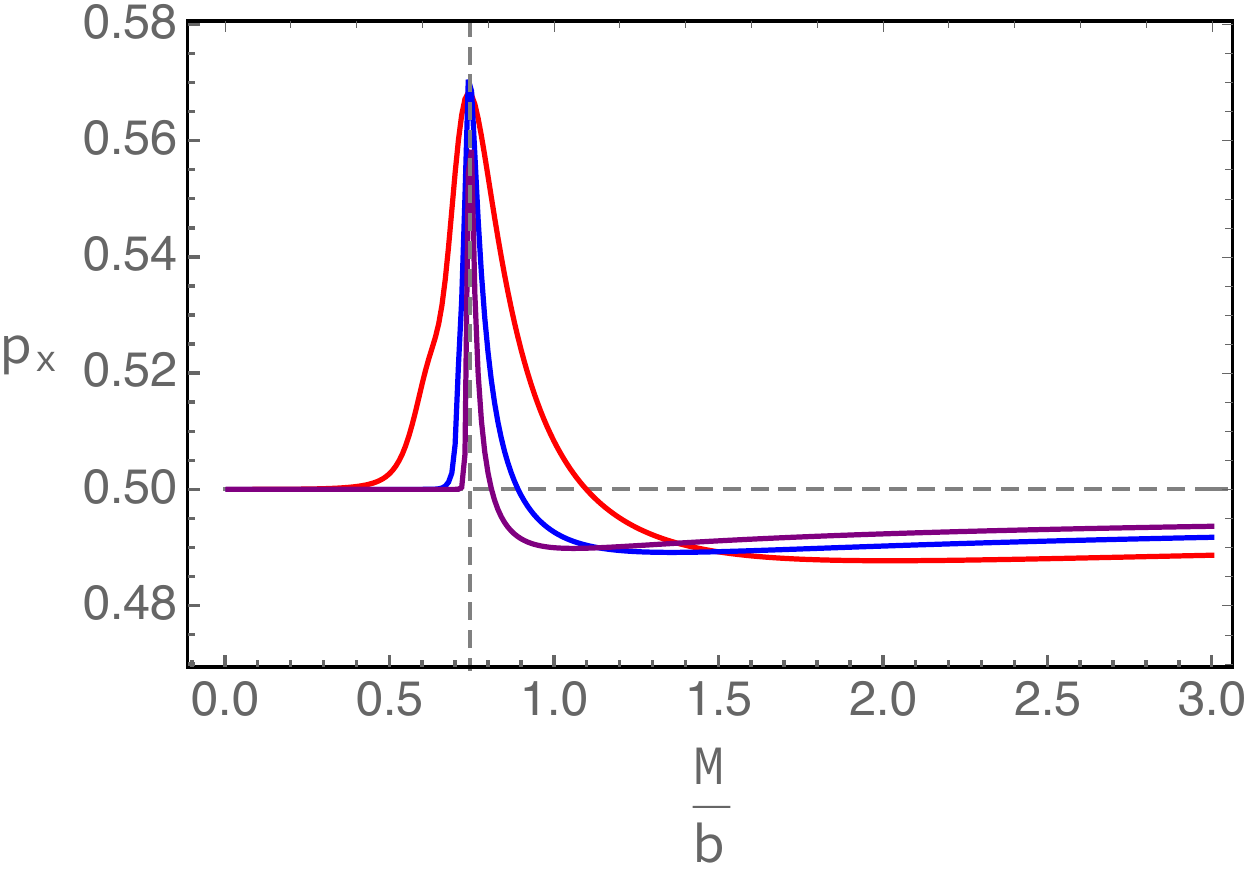}~~
\includegraphics[width=0.444\textwidth]{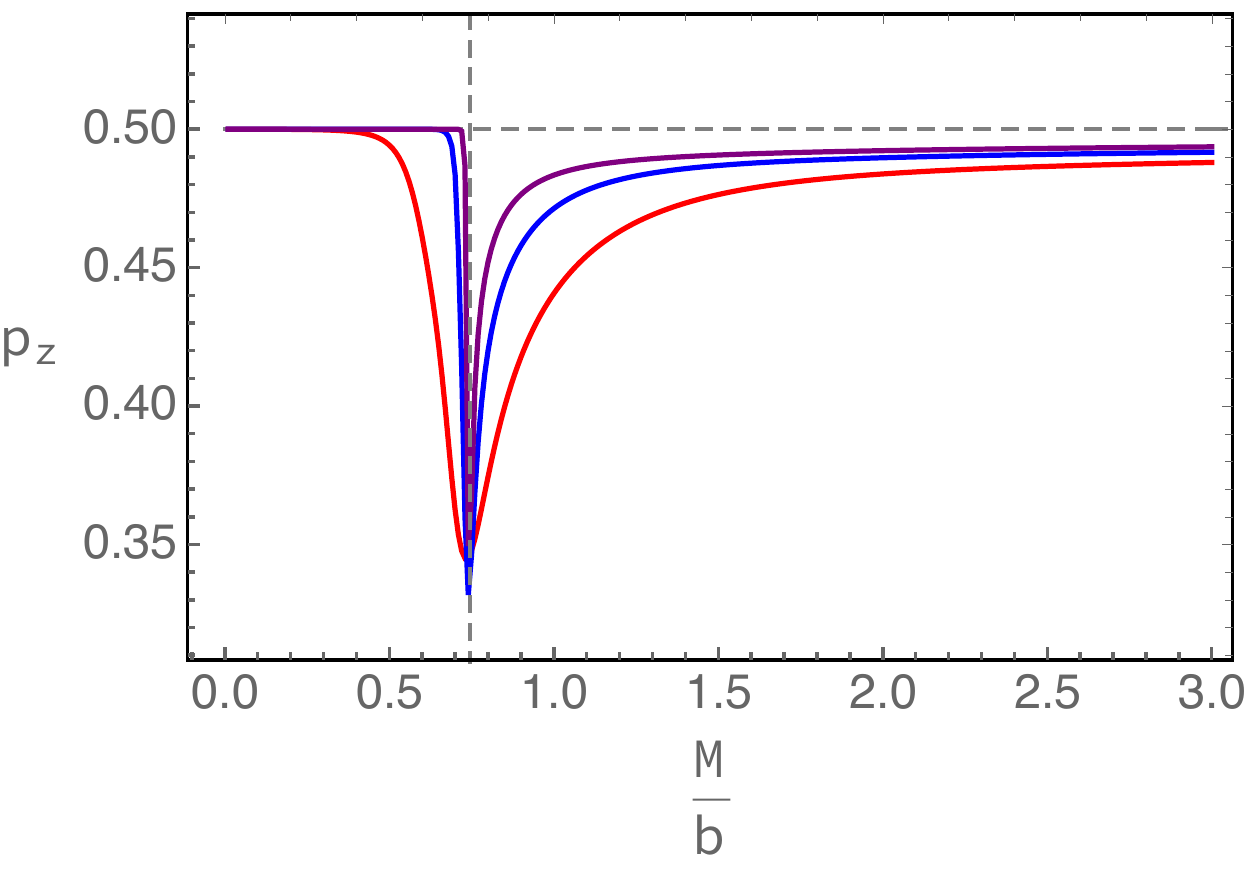}
\end{center}
\vspace{-0.3cm}
\caption{\small  Plots of Kasner exponents as a  function of $M/b$. For all cases we have $T/b=0.05$ (red), $0.02$ (blue), $0.01$ (purple). The dashed gray vertical lines are the Kasner exponents of five dimension Schwarzschild black hole.}
\label{fig:wsmkas}
\end{figure}

In Fig. \ref{fig:wsmkas}, we show the Kasner exponents as functions of $M/b$ at different temperatures $T/b=0.05$ (red), $0.02$ (blue), $0.01$ (purple). We find that at low temperature, the Kasner exponents in the Weyl semimetal phase take the same value as for the Schwarzschild black hole (e.g. within the difference of order less than $10^{-9}$ between $M/b=0.5$ and $M/b=0$ at $T/b=0.01$). This reminds us the topological feature of the topological phase in terms of the black hole singularity.  It is related to the fact that the matter fields do not backreact relevantly to the Schwarzschild solution in the topological phase, i.e. the probe limit of system in terms of matter fields in the  Schwarzschild black hole background works well. We have also checked that inside the black holes, in the topological phase the matter fields obtained from the backreacted case  match well with the solutions obtained from the probe limit.
In the quantum critical regime, the Kasner exponents 
is not monotonic. 
While in the trivial phase, the Kasner exponents behave similarly to those of the static hairy black holes in the Einstein-scalar theory \cite{Frenkel:2020ysx}. 

Note that in \cite{Liu:2018bye}, a paradigm for constructing the topological phase was proposed and the holographic Weyl semimetal belongs to the first type, where the matter fields are irrelevant in the IR of the Schwarzschild black hole. It seems likely that in any topological phase of this kind, the singularities are of Kasner form taking values of Schwarzschild black hole.

\subsection{Proper time of timelike geodesics}

One interesting connection between the interior geometry and the boundary observable is given in  \cite{Grinberg:2020fdj} that the proper time of radial timelike geodesic can be encoded in the thermal one point functions of heavy operators. Therefore it is interesting to study the proper time of radial timelike geodesics to see if it has specific behavior during the topological phase transitions. 

\begin{figure}[h!]
\begin{center}
\includegraphics[width=0.47\textwidth]{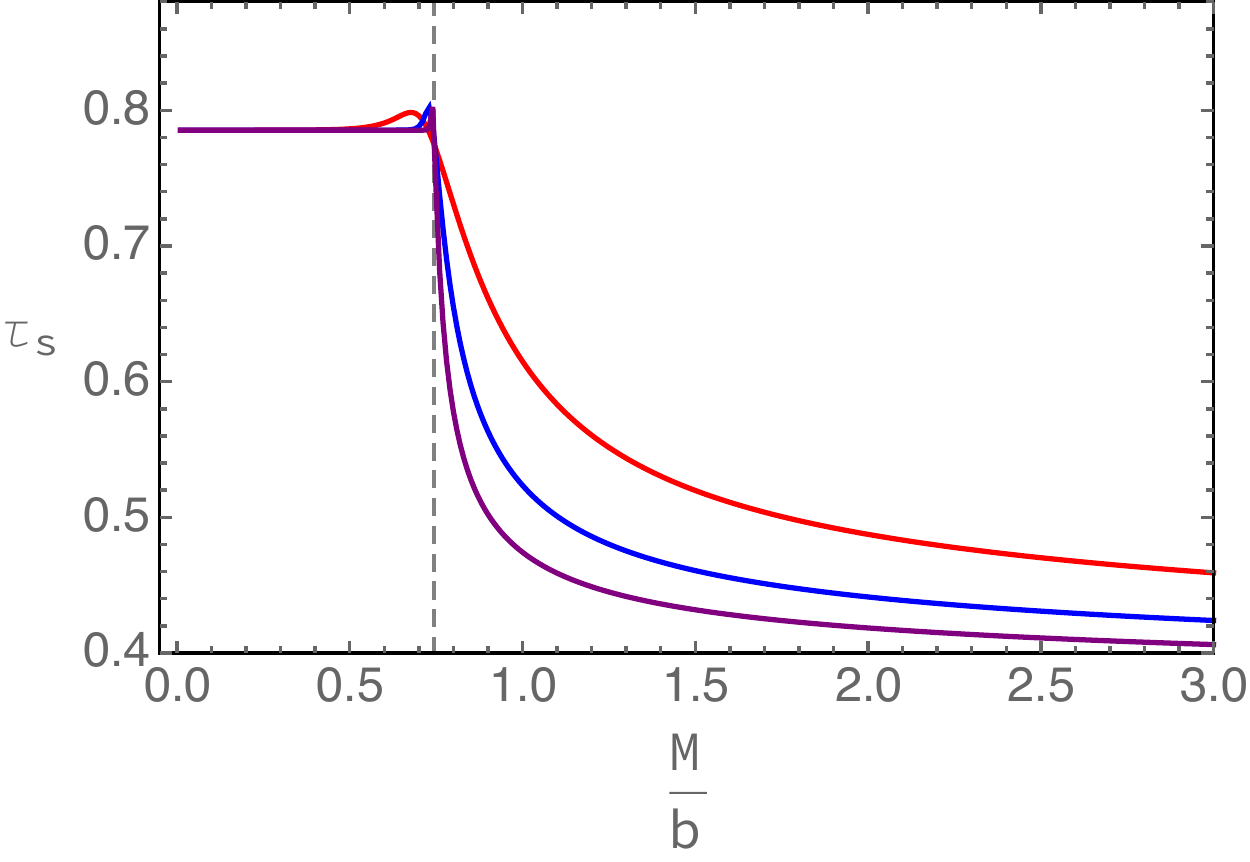}
\end{center}
\vspace{-0.3cm}
\caption{\small  Plots of the proper time $\tau_s$ from the horizon to the singularity as a function of $M/b$ at different temperatures $T/b=0.05$ (red), $0.02$ (blue), $0.01$ (purple).}
\label{fig:taus}
\end{figure}

We consider radial timelike geodesic for which $g_{tt} {\dot t}^2+ g_{rr} {\dot r}^2 = -1$, where the dot denotes the derivative with respect to the proper time $\tau$.  
Along the geodesic there is a conserved charge $E = - g_{tt} \dot t$ which can be interpreted as energy. 
Then the equation of motion of the geodesic becomes 
\begin{align}
    \frac{E^2}{g_{tt}}+ g_{rr} {\dot r}^2 = -1\,,
\end{align}
from which we obtain
\begin{align}
    \frac{d\tau}{dr} = \frac{1}{\sqrt{E^2-u}}\,.
\end{align}
The proper time from the horizon to the singularity of a particle with $E=0$ (i.e. the longest time) is
\begin{align}
    \tau_s= \int_{r_s}^{r_h} \frac{dr}{\sqrt{-u}} \,.
\end{align}
The plots of $\tau_s$ as a function of $M/b$ for different $T/b$ are shown in Fig. \ref{fig:taus}.

The proper time from the horizon to the singularity in the topological phase is equal to the case of Schwarzschild black hole $\tau_s= \pi/4$ (e.g. within the difference of order less than $10^{-4}$  between $M/b=0.5$ and $M/b=0$ at $T/b=0.01$). This is expected from the fact that in the topological phase at low temperature the interior of the black holes match well with the Schwarzschild black hole. In the topologically  trivial phase $\tau_s$ is monotonically decreasing. Moreover, $\tau_s$ shows a jump behavior and takes a maximum value in the critical regime. 
Note that $\tau_s$ is encoded in the thermal one point function of heavy operators in the form of $\langle\mathcal{O}\rangle\propto e^{-im\tau_s} $ where the complexified mass $m$ has $\text{Im}(m)<0$ \cite{Grinberg:2020fdj}. One might use this thermal one point function as the ``order" parameter for the topological  phase transition.  
The behavior of the proper time also reminds us  the behavior of the dimensionless information screening length in \cite{Baggioli:2018afg}. One obvious  difference is that the information screening length is determined by the geometric quantities at the horizon, while $\tau_s$ is determined by the inner  geometry from the horizon to the singularity.

\section{Inside holographic nodal line semimetal}
\label{sec3}

In the previous section, we have seen that the interior of black hole geometries for the holographic WSM exhibit interesting behavior. In the topological WSM phase, the Kasner exponents of the dual geometries take the same value as for the Schwarzschild black hole at low temperature, as shown in Fig. \ref{fig:wsmkas}. Moreover, the dual operator which encodes the proper time from the horizon to the singularity could serve as an ``order parameter" during the topological phase transition, as shown in Fig. \ref{fig:taus}. To check if these behaviors are universal for any topological phase transitions, in this section we study the other topological phase transition model from holography, i.e. the holographic NLSM model which describes a phase transition from the topological NLSM phase to a trivial semimetal phase \cite{Liu:2018bye, Liu:2020ymx}.  

The action for the holographic NLSM \cite{Liu:2020ymx} is 
\begin{align}
\label{eq:actionnlsm}
\begin{split}
     S&= \int d^5x \,\sqrt{-g}\, \bigg[ \frac{1}{2\kappa^2}\big(R+\frac{12}{L^2}\big) -\frac{1}{4}\mathcal{F}^2 -\frac{1}{4}F^2 +\frac{\alpha}{3} \epsilon^{abcde} A_{a} \big( F_{bc}F_{de} +3 \mathcal{F}_{bc} \mathcal{F}_{de} \big) \\
   &- (D_{a}\Phi)^*(D^{a}\Phi)-V_1(\Phi) - \frac{1}{6\eta} \epsilon^{abcde} \big( i B_{ab} H^*_{cde} - i B^*_{ab} H_{cde} \big) -V_2(B_{ab}) - \lambda |\Phi|^2 B^*_{ab}B^{ab} \bigg]\,,
\end{split}
\end{align}
where $\mathcal{F}_{ab} = \partial_a V_b-  \partial_b V_a$ is the vector gauge field strength. $F_{ab} = \partial_a A_b - \partial_b A_a$ is the axial gauge field strength. 
$D_a = \nabla_a - i q_1 A_a$ is the covariant derivative and $q_1$ is the axial charge of scalar field. $\alpha$ is the Chern-Simons coupling. $B_{ab}$ is an antisymmetric complex two form field with the field strength
\begin{align}
H_{abc} = \partial_a B_{bc} + \partial_b B_{ca} + \partial_c B_{ab} - i q_2 A_a B_{bc} - i q_2 A_b B_{ca} - i q_2 A_c B_{ab}\,,
\end{align}
where $q_2$ is the axial charge of the two form field. $\eta$ is the Chern-Simons coupling strength of the two form field. The introduction of the Chern-Simons terms (i.e. the $B\wedge H^*$ term) while not a canonical kinetic term for the two form field (i.e. the $H_{abc}H^{abc*}$ term) follows from the self-duality condition (i.e. $\bar{\psi}\gamma^{\mu\nu}\gamma^5\psi=-
\frac{i}{2}
\varepsilon^{\mu\nu}_{~~~\alpha\beta}\bar{\psi}\gamma^{\alpha\beta}\psi)$ of the anti-symmetric tensor operators in the weakly coupled theory \cite{Liu:2020ymx}.  
The potential terms in \eqref{eq:actionnlsm} are chosen as
\begin{align}
V_1 = m_1^2 |\Phi|^2 + \frac{\lambda_1}{2} |\Phi|^4\,,\qquad V_2 = m_2^2 B_{ab}^* B^{ab}\,, 
\end{align}
where $m^2_1$ and $m_2^2$ are the mass parameters of the scalar field and the two form field. The $\lambda$ term in the action \eqref{eq:actionnlsm} denotes the interaction between the scalar field and the two form field. We set $2\kappa^2=L=1$.

Similar to the holographic WSM, we focus on the finite temperature solution and take the ansatz
\begin{align}
\label{eq:nlansatz}
    \begin{split}
        ds^2 &= -udt^2+\frac{dr^2}{u}+ f(dx^2+dy^2)+hdz^2\,,\\
        \Phi&=\phi\,,\\
        B_{xy}&=-B_{yx}=\mathcal{B}_{xy}\,,\\
        B_{tz}&=-B_{zt}=i \mathcal{B}_{tz}\,.
    \end{split}
\end{align}
Plugging the above ansatz into the equations of motion, we obtain the dynamical equations of the fields, which can be found in the appendix \ref{app:nlsm}. In the following we choose $m_1^2=-3, m_2^2=1, \eta=2$ and $q_1=q_2=1, \lambda=1, \lambda_1=0.1$ for simplicity. 

With the following boundary conditions, 
\be
\lim_{r\to\infty}r\phi=M\,,~~~
\lim_{r\to\infty}
\frac{\mathcal{B}_{xy}}{r}=\lim_{r\to\infty}
\frac{\mathcal{B}_{tz}}{r}=b\,,
\ee
we can integrate the system from the boundary to the horizon. Different from the holographic WSM, in the holographic NLSM there is no sharp ``order parameter" like anomalous Hall conductivity.  Nevertheless, it was found in \cite{Liu:2020ymx} that at zero temperature, the dual fermionic spectral function shows multiple Fermi surfaces with the topology of nodal lines when $M/b<(M/b)_c$ while it is gapped when $M/b>(M/b)_c$. This indicates that the system undergoes a topological phase transition from a topological NLSM phase to a topologically trivial semimetal phase.  

With the regularity condition near the horizon, the system can be further integrated to the singularity. In the following we will discuss the interior geometries and singularities of the system. 

\subsection{Behaviors of Kasner exponents}

Close to the singularity $r\to r_s$, similar to the holographic WSM case we again take the ansatz 
\begin{align}
\label{eq:asns2}
    u\sim -u_0 (r-r_s)^{n_u}\,,~~~ f\sim f_0 (r-r_s)^{n_f}\,,~~~ h\sim h_0 (r-r_s)^{n_h}\,,~~~ \phi\sim n_{\phi} \ln (r-r_s)\,,
\end{align}
where $u_0, f_0, h_0$ and $n_u, n_f, n_h, n_\phi$ are all constants. The other two matter fields $\mathcal{B}_{tz}$ and  $\mathcal{B}_{xy}$ will be determined by the above ansatz. 

The equations of motion can be simplified close to the singularity under the assumption that the ignored terms are subleading 
\begin{align}
\label{eq:nhnlsm}
\begin{split}
    \frac{u''}{u} - \frac{f''}{f} + \frac{h'}{2h} \left( \frac{u'}{u} - \frac{f'}{f}  \right)&=0\,,\\
     \frac{u''}{2u}+\frac{f''}{f}-\frac{f'^2}{4f^2} + \frac{f'u'}{fu}+\frac{1}{2} \phi'^2 &=0\,,\\
    \frac{f'^2}{4f^2} +\frac{f'h'}{2fh} + \frac{u'}{2u}\left( \frac{f'}{f} + \frac{h'}{2h} \right)-\frac{1}{2}\phi'^2 &=0\,,\\
    \phi'' + \big( \frac{f'}{f}+\frac{h'}{2h}+ \frac{u'}{u} \big)\phi'&=0\,,\\
    \mathcal{B}_{tz}'- \frac{\eta \sqrt{h}}{2f}( \lambda \phi^2) \mathcal{B}_{xy} &=0\,,\\
    \mathcal{B}_{xy}'- \frac{\eta f}{2\sqrt{h} u}( \lambda \phi^2) \mathcal{B}_{tz} &=0\,.
\end{split}
\end{align}
The first four equations in \eqref{eq:nhnlsm} are the same as the ones in the holographic WSM. Similarly, we obtain
\begin{align}
n_h=2\,(1-n_u-n_f)\,, ~~~~~
n_{\phi}=\pm \sqrt{(2n_f+n_u)(1-n_u)- \frac{3n_f^2}{2}} \,.
\end{align}
From the last two equations in \eqref{eq:nhnlsm} we have the following leading order solutions for the two form fields near the singularity 
\be
\label{eq:2formnh}
\mathcal{B}_{xy}\sim \mathcal{B}_{xy0}+\dots\,,~~~~
\mathcal{B}_{tz}\sim \mathcal{B}_{tz0}+\dots\,,~~~~
\ee
where the dots are subleading terms with the dominate terms of form $(r-r_s)^{2-n_u-2 n_f}(\log (r-r_s))^2$ and $(r-r_s)^{2n_f} (\log (r-r_s))^2$ respectively. Here we have assumed $2-n_u-2n_f>0$ and $n_f>0$, otherwise the leading solution of the two form field might be divergent. Similar to the holographic WSM, these  constants of the two form field depend on the scaling symmetry of the system. 

Note that in \eqref{eq:nhnlsm} we have assumed that the ignored terms are subleading. More explicitly, we have assumed 
\be
\label{eq:hnls-paregime}
    n_u<2\,,~~~2n_u+n_h<2\,, ~~~
    n_u+2n_f<2\,. 
\ee
Note that the last two inequalities of above are consistent with the assumptions used in obtaining  \eqref{eq:2formnh}. 
We have checked numerically that the inequalities \eqref{eq:hnls-paregime} are satisfied for the parameters we have considered. 

Similar to the discussion in section \ref{subsec:wsmkas}, we can make 
 a coordinate transformation \eqref{eq:coortr} to write the metric \eqref{eq:asns2} into the Kasner form as \eqref{eq:kasnerform} with the parameters \eqref{eq:ptnurel} and the Kasner relations \eqref{eq:kasrel}. Here the leading order of the two form fields are constant close to the singularity. 

The two conserved charges of the scaling symmetries are
\bea
    Q_1&=& \frac{8}{\eta} \mathcal{B}_{tz} \mathcal{B}_{xy} + \frac{u}{\sqrt{h}} (f'h-fh')\,,\\
    Q_2&=& \frac{f}{\sqrt{h}} (u'h-uh')\,.
\eea
Evaluating them at the horizon and at the singularity we obtain
\be
\frac{8}{\eta}\mathcal{B}_{xy0}\mathcal{B}_{tz0}=u_0f_0\sqrt{h_0}(n_f-n_h)
\ee
and
\begin{align}
   4\pi T f_1 \sqrt{h_1} =Ts= u_0 f_0 \sqrt{h_0} (2-2 n_f-3 n_u)
\end{align}
where $s$ is the density of entropy. We have checked the above relations numerically. 

\begin{figure}[h!]
\begin{center}
\includegraphics[width=0.444\textwidth]{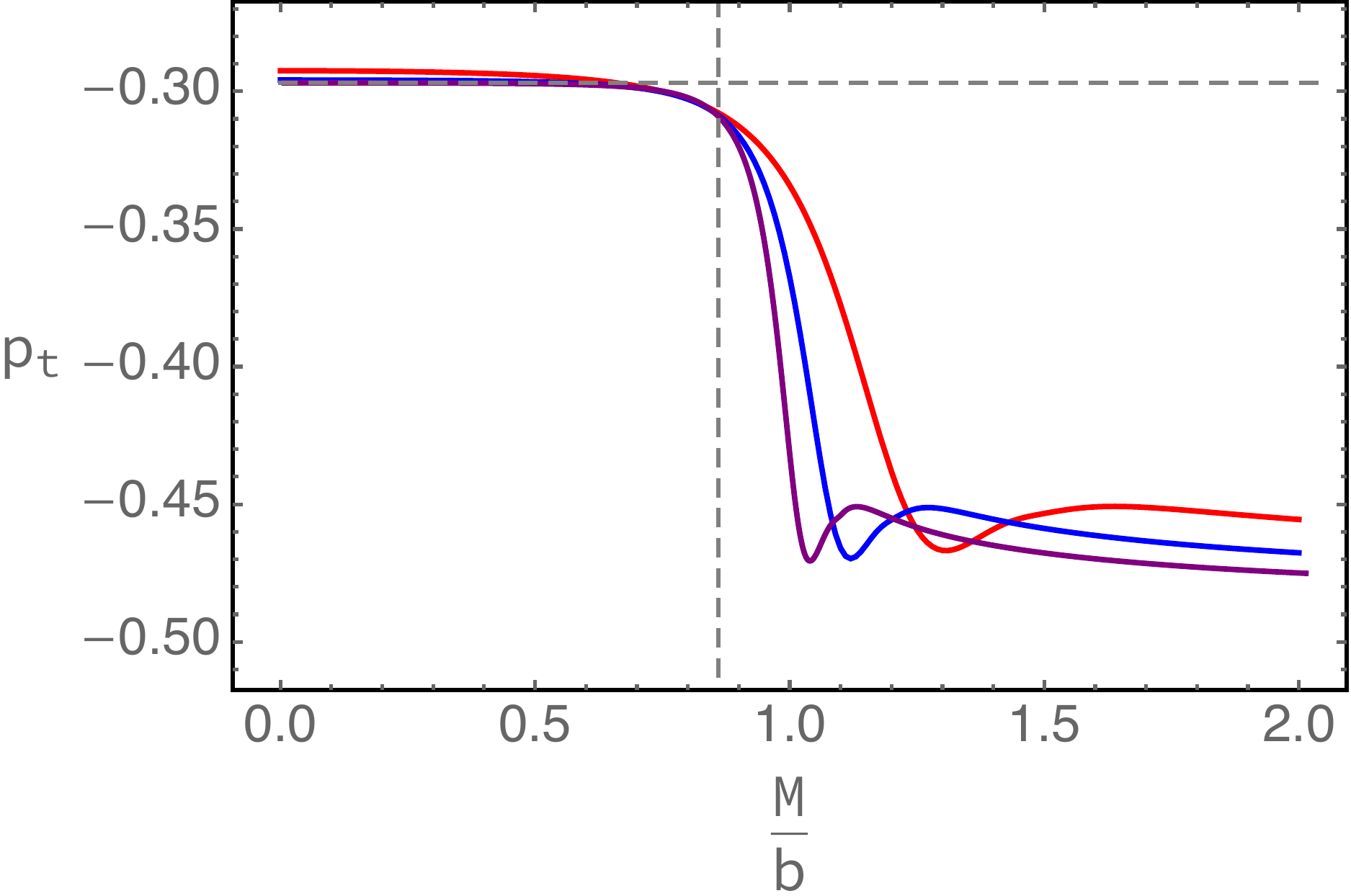}~~
\includegraphics[width=0.444\textwidth]{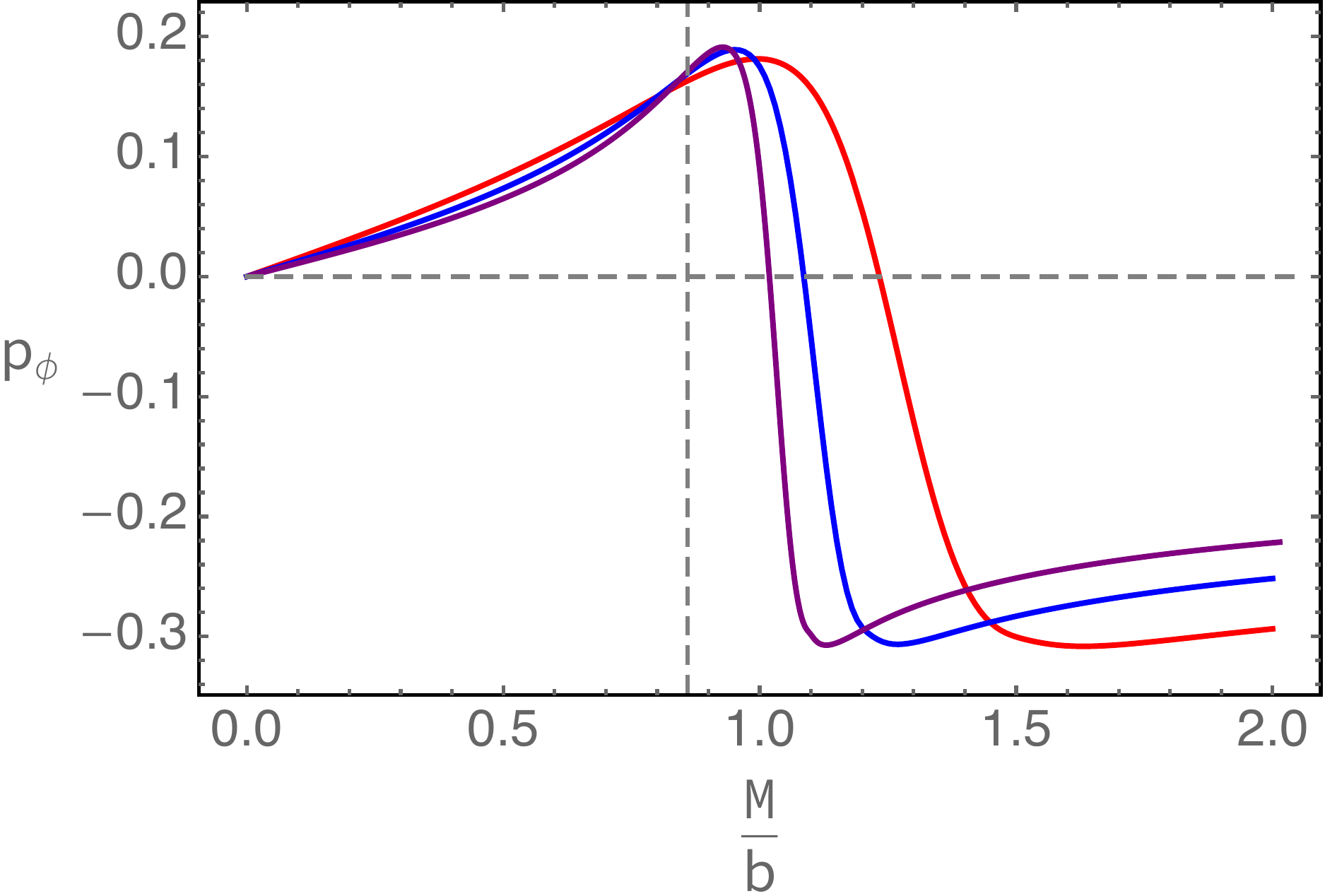}\\
\includegraphics[width=0.444\textwidth]{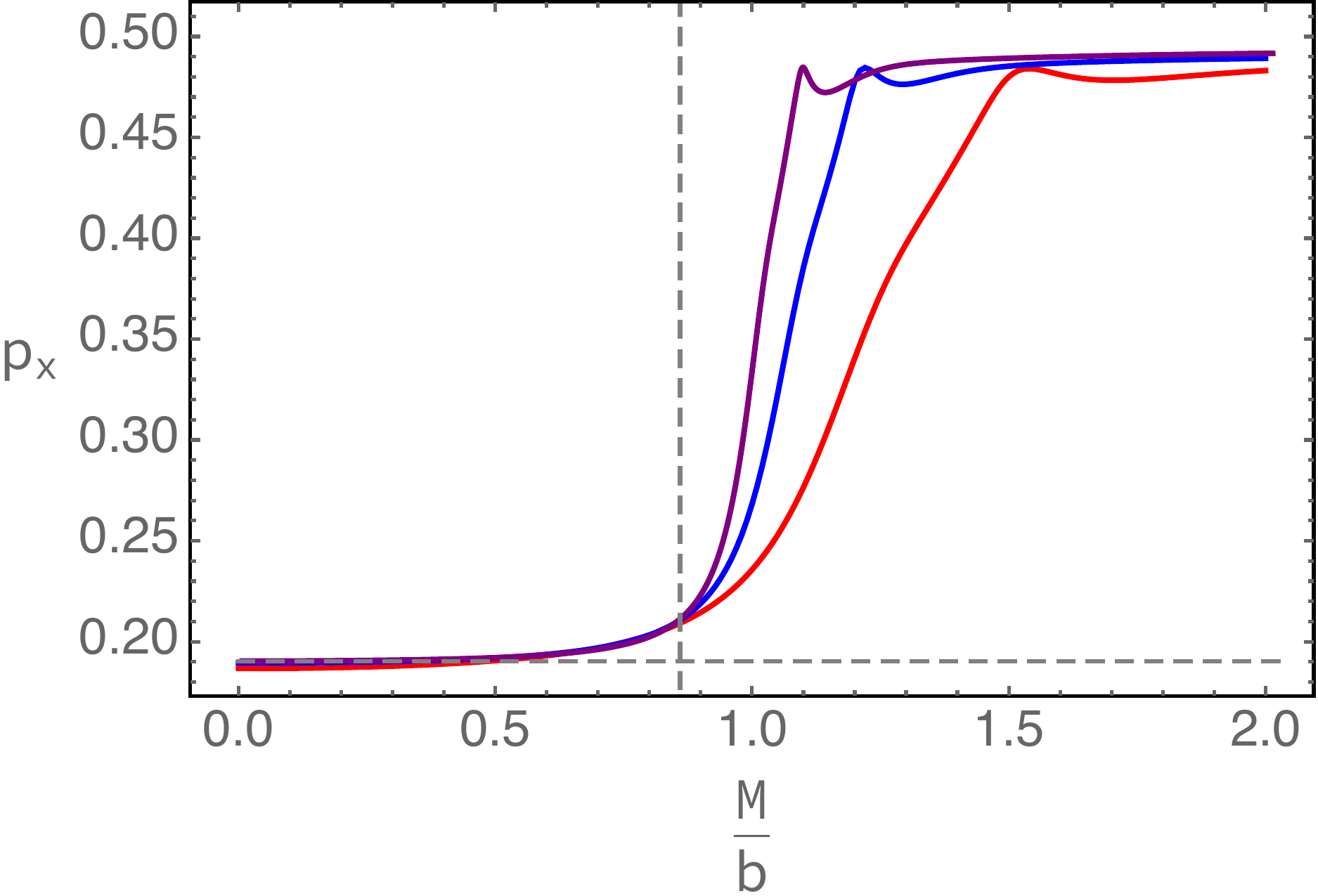}~~
\includegraphics[width=0.444\textwidth]{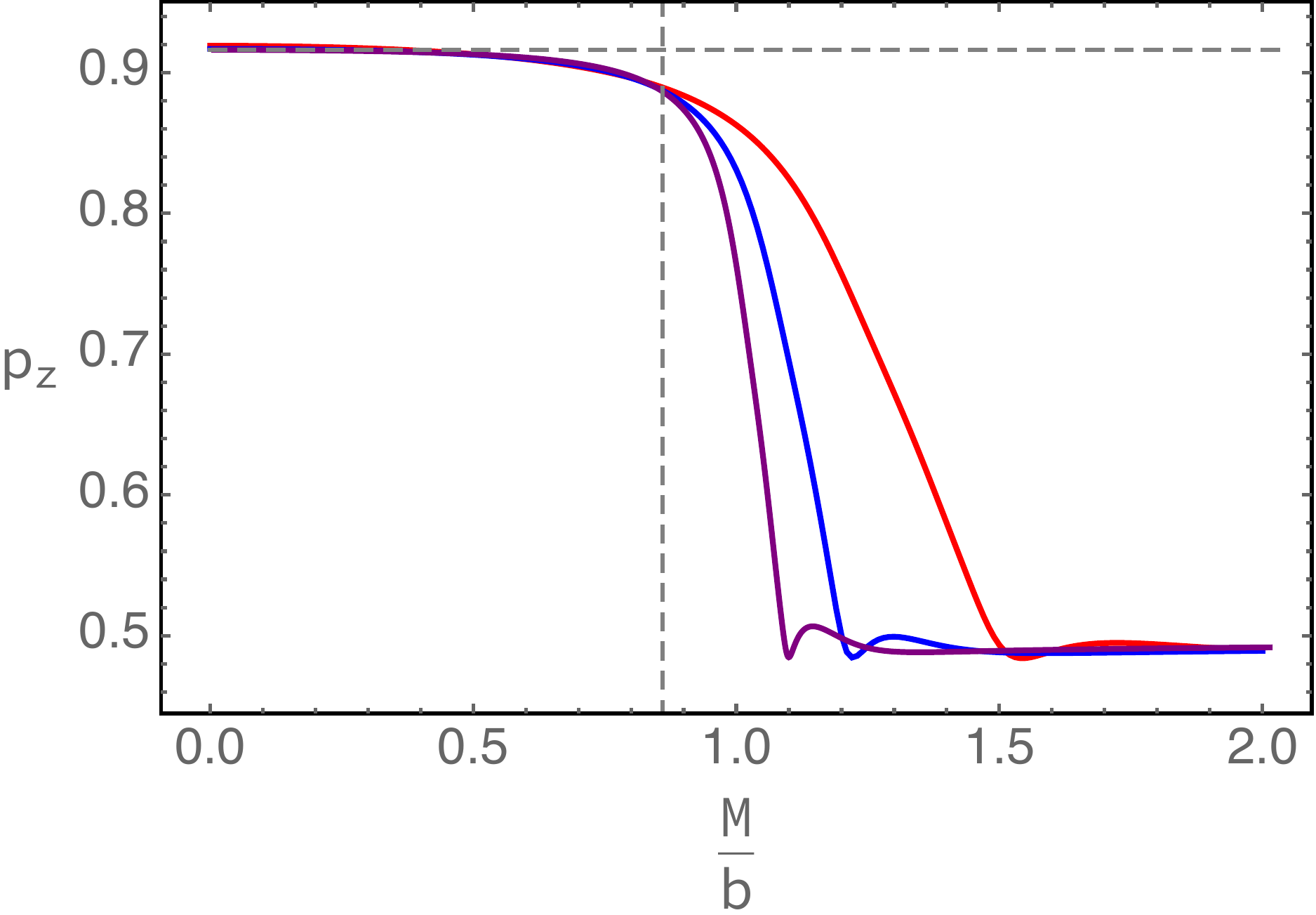}
\end{center}
\vspace{-0.3cm}
\caption{\small  Plots of Kasner exponents for holographic NLSM as functions of $M/b$. For all cases we have $T/b=0.05$ (red), $0.02$ (blue), $0.01$ (purple). The horizontal dashed gray lines represent the Kasner exponents for $M/b=0$ at $T/b=0.01$. The vertical dashed gray lines represent the quantum critical point at zero temperature.}
\label{fig:nlsmkas}
\end{figure}

In Fig. \ref{fig:nlsmkas}, we show the Kasner exponents for the holographic NLSM as functions of $M/b$ at different temperature $T/b=0.05$ (red), $0.02$ (blue), $0.01$ (purple). We find that at low temperature, the Kasner exponents $p_t, p_x, p_z$ of the metric fields in the NLSM semimetal phase are almost constant as functions of $M/b$ in the topological phase (e.g. within the difference of order less than $1\%$ 
between $M/b=0.5$ and $M/b=0$ at $T/b=0.01$), which is quite similar to the holographic WSM,  while $p_\phi$ as a function of $M/b$ changes a lot in the topological phase. Note that this is consistent with the Kasner relations \eqref{eq:kasrel} since $p_\phi$ is small. It is expected that at extremely low temperature, the properties of the Kasner exponents in the holographic NLSM might be the same as those in the holographic WSM, i.e. all the Kasner exponents are constant as functions of $M/b$. Due to numerical difficulty we have not explored such a low temperature regime.  

\begin{figure}[h!]
\begin{center}
\includegraphics[width=0.244\textwidth]{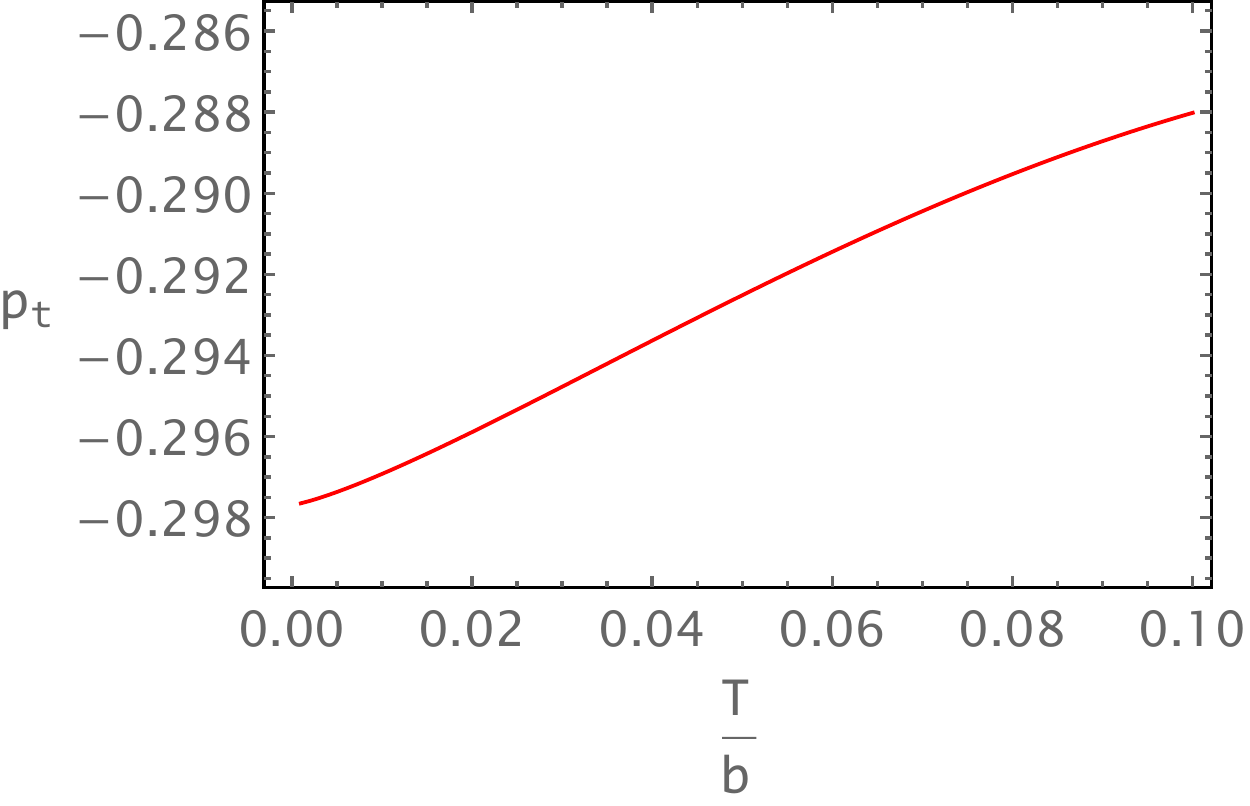}
\includegraphics[width=0.244\textwidth]{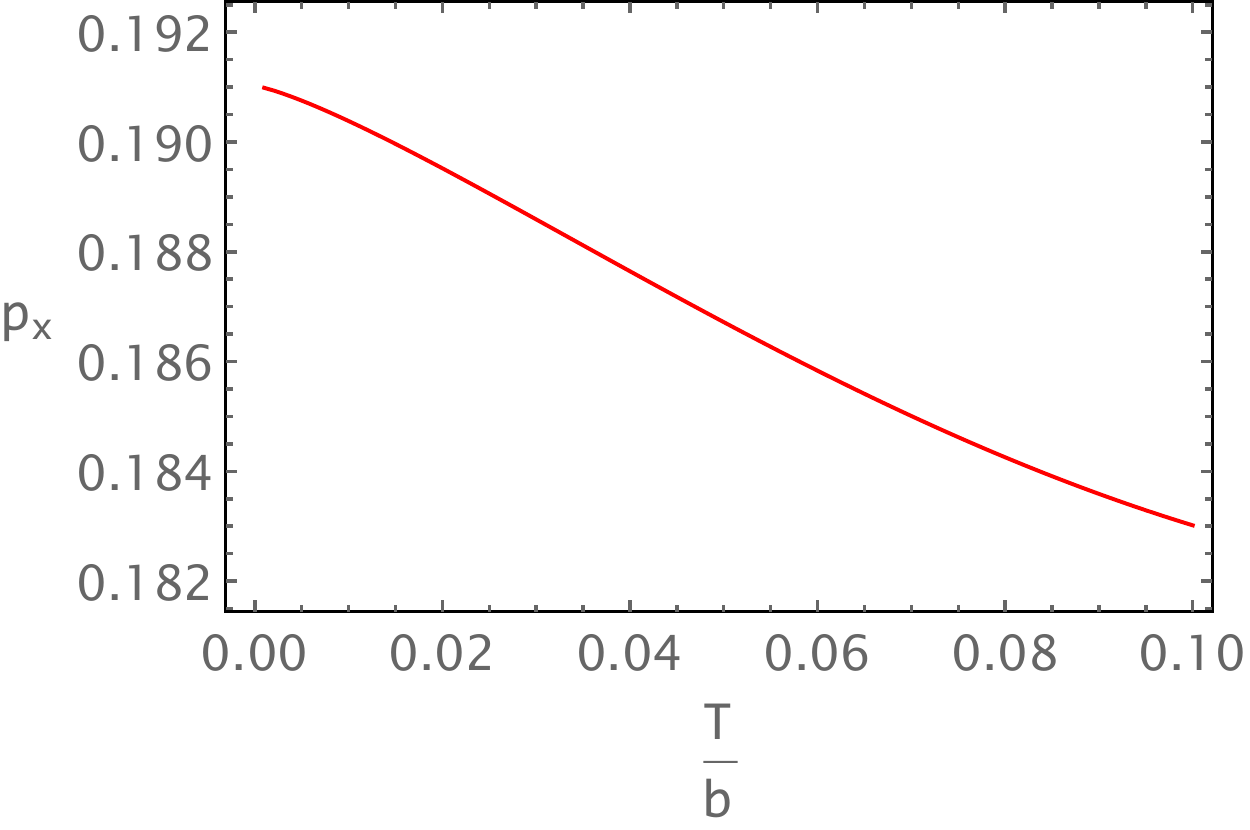}
\includegraphics[width=0.244\textwidth]{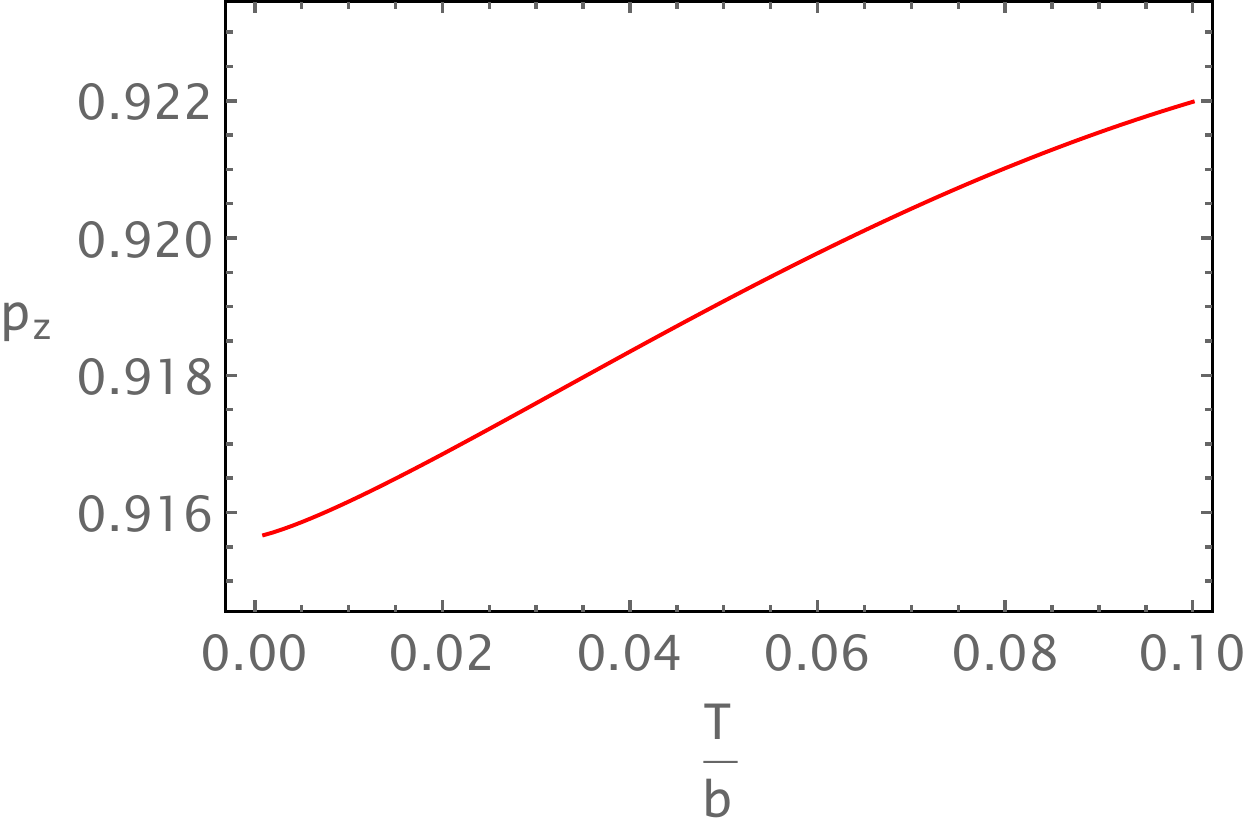}
\includegraphics[width=0.244\textwidth]{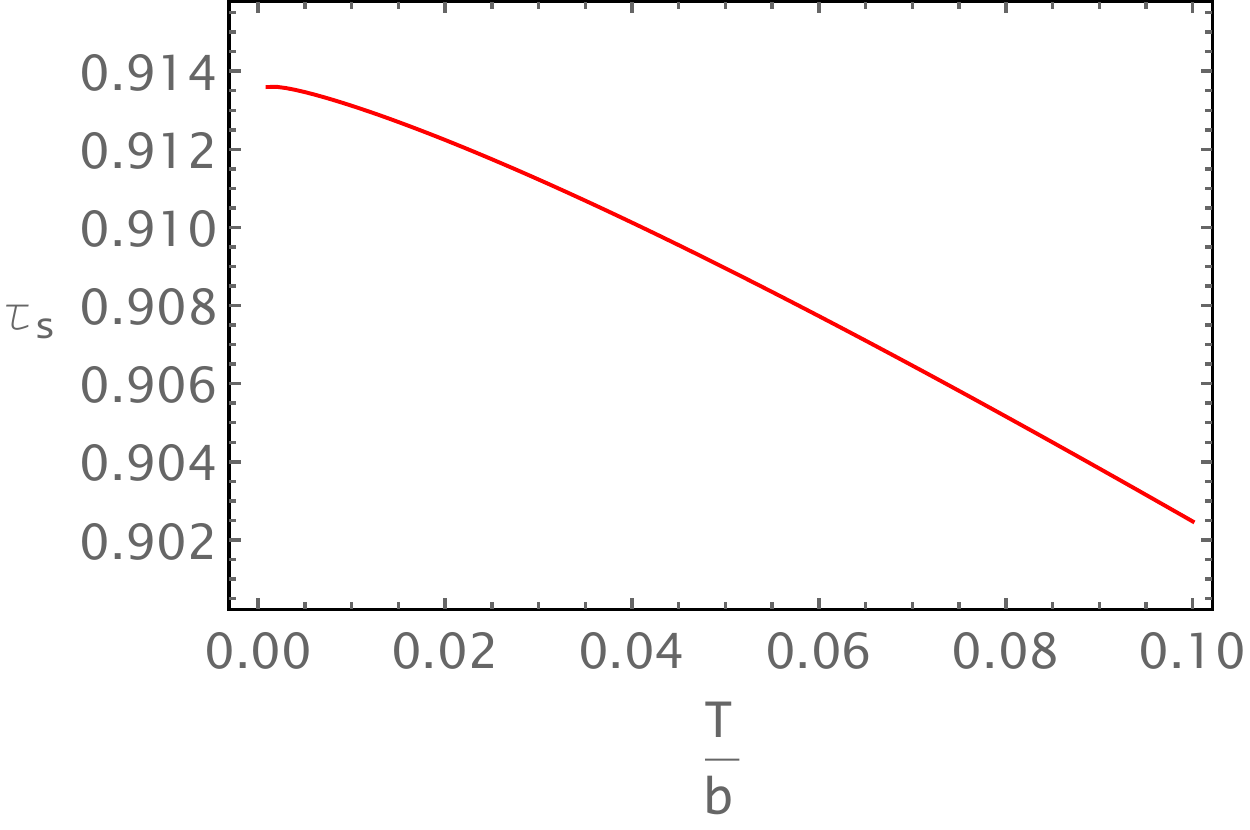}
\end{center}
\vspace{-0.4cm}
\caption{\small  Plots of Kasner exponents and $\tau_s$ for holographic NLSM as functions of $T/b$ when $M/b=0$. }
\label{fig:topNLSMkas}
\end{figure}

Different from the holographic WSM where the geometry is the same as Schwarzschild black hole with a constant nonzero $A_z$ when $M/b=0$. Here when $M/b=0$, in the holographic NLSM, due to the fact that the matter fields strongly backreact to the IR geometry of Schwarzschild black hole, the Kasner exponents are no longer the constant exponents of Schwarzschild black hole. Instead these Kasner exponents depend on $T/b$, as shown in the first three pictures in Fig. \ref{fig:topNLSMkas}. Nevertheless, at low enough temperature the Kasner exponents are also nearly constant as functions of $T/b$. Combining with the results from Fig. \ref{fig:nlsmkas}, we conclude that in the topological NLSM phase the Kasner exponents of the metric fields are almost constant as functions of $M/b$ and $T/b$ at extremely low temperature.

\vspace{-0.1cm}

\subsection{Proper time of timelike geodesics}

\begin{figure}[h!]
\begin{center}
\includegraphics[width=0.47\textwidth]{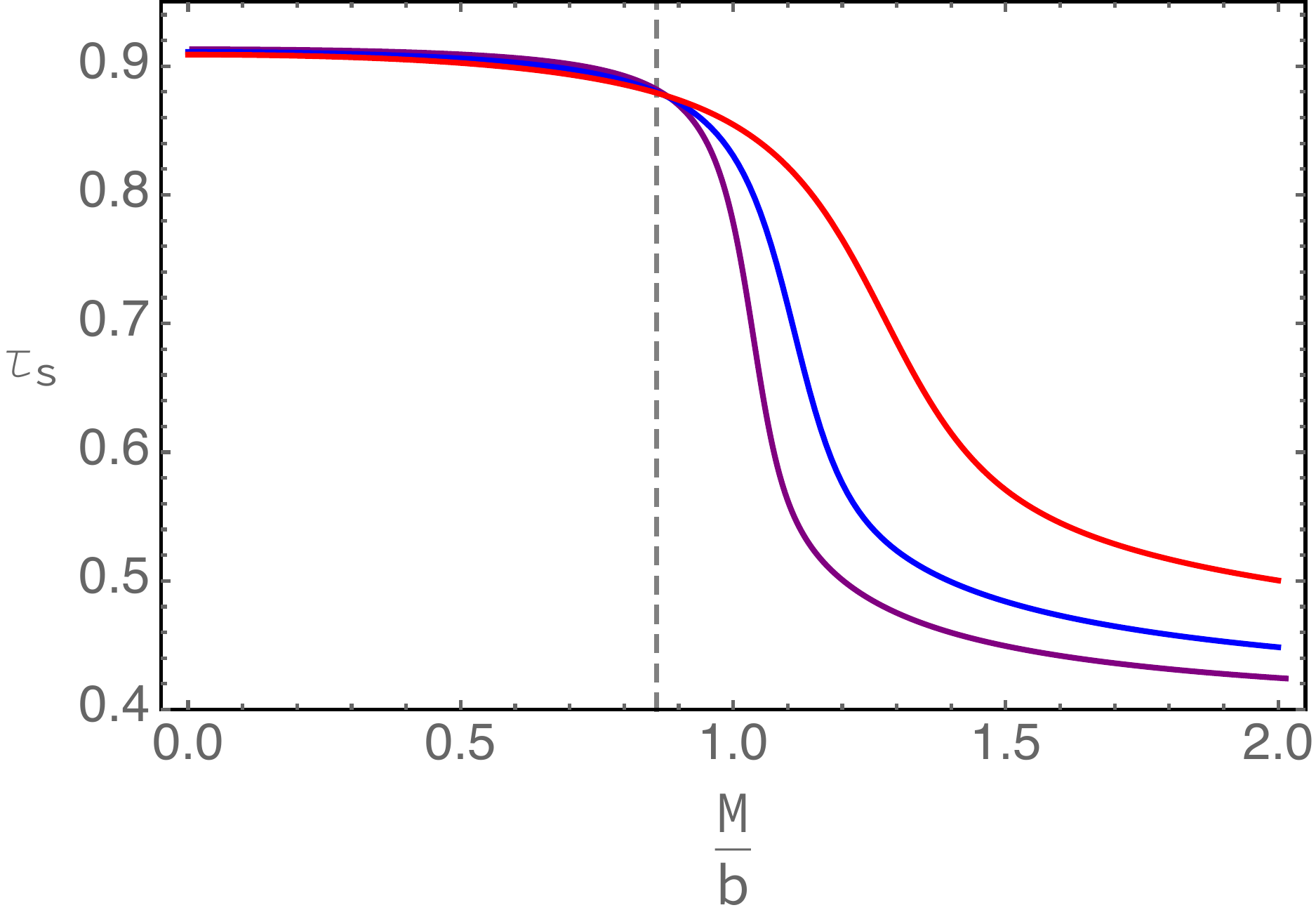}
\end{center}
\vspace{-0.4cm}
\caption{\small  Plots of the proper time $\tau_s$ from the horizon to the singularity as a function of $M/b$ at different temperatures $T/b=0.05$ (red), $0.02$ (blue), $0.01$ (purple).}
\label{fig:nlsmtaus}
\end{figure}

Similar to the holographic WSM, we can also discuss the proper time from the horizon to the singularity in holographic NLSM. In Fig. \ref{fig:nlsmtaus}, we show the proper time $\tau_s$ as a function of $M/b$ at different temperatures. Again we see that at low temperature, the proper time as a function of $M/b$ and $T/b$ is almost a constant in the topological phase  (e.g. within the difference of order less than $5\text{\textperthousand}$  
between $M/b=0.5$ and $M/b=0$ at $T/b=0.01$), which shows a topological behavior under the changes of the systems. Similar to the holographic WSM, we could take the operator which encodes the information of $\tau_s$ as the order parameter for the topological phase transition in the holographic NLSM. In the trivial phase, the proper time $\tau_s$ is monotonically decreasing when we increase $M/b$. 

\section{Conclusion and discussion}
\label{sec4}

We have studied the interior geometries of black holes in two different holographic topological semimetals. We find that the singularities of the geometries are of simple Kasner form, together with a constant one form gauge potential or constant two form tensor fields. In the topological WSM phase, all the Kasner exponents are constant taking values of Schwarzschild black hole at low temperature. In the topological NLSM phase, the Kasner exponents of the metric fields are also almost constant as  functions of $M/b$ and $T/b$ at extremally low temperature 
while the Kasner exponent of the scalar field is small and changes a bit in the topological phase. Moreover, we find the proper time from the horizon to the singularity is nearly constant as a function  of $M/b$ and $T/b$ in both holographic WSM and holographic NLSM. These features seem to be of topological in the sense that they stay as constant during the changes of physical parameters of the systems. The proper time in the trivial phases of the two holographic semimetal decreases when we increase $M/b$.
 
In addition to the above universal behavior, specific behaviors inside the horizon are also found. In the topological phase of holographic WSM, we find the oscillations of the matter field $\phi$ inside the horizon at low temperature. In other phases we have not found any oscillations of fields. The Kasner exponents 
are not monotonic 
in the critical regime of holographic WSM. There is no oscillation of background fields in holographic NLSM. In the trivial phases of the two holographic semimetals, the Kasner exponents behave differently, where the details can be found in Fig. \ref{fig:wsmkas} and Fig. \ref{fig:nlsmkas}.   

It is known that for holographic topological semimetals at zero temperature, the bulk geometries in IR have different Lifshitz scaling exponents for different phases. We should emphasize that the Kasner exponents, the proper time from the horizon to the singularity and the  near horizon quantum Lifshitz scaling exponents are of different geometric aspects in the bulk. The first and second physical quantities are from the black hole interior at finite temperature while the last quantity is from the geometry near the horizon at zero temperature. There is no clear connection among them. The quantum Lifshitz scaling exponents near the horizon can be manifested in terms of transports at low temperature, e.g, the dependence of the conductivity on temperature or frequency in holographic WSM \cite{Grignani:2016wyz} and holographic NLSM \cite{Rodgers:2021azg}, while the first two qunatities can be extracted from the correlators of heavy operators. 

It is very interesting to determine the precise observables associated to the Kasner exponents and the proper time to understand the role played by topology. Perhaps an explicit top-down holographic topological semimetal would be helpful. It would be interesting to connect the topological features of Kasner exponents and the proper times in the topological phases of the two holographic semimetals to the topological invariants.  This would shed light on the universal theories describing the topological semimetals. Meanwhile, it is also interesting to check the behavior of these physical quantities in the topological phases of other holographic topological semimetals, e.g. \cite{Juricic:2020sgg, Ji:2021aan}, to check if they are universal feature of topological semimetals. 

\vspace{.8cm}
\subsection*{Acknowledgments}
 We are grateful to Matteo Baggioli, Karl Landsteiner, Ya-Wen Sun, Xin-Meng Wu, Jun-Kun Zhao for useful discussions. This work is supported by the National Natural Science Foundation of China grant No.11875083. 

\vspace{.6 cm}
\appendix
\section{Equations in holographic WSM}
\label{app:hwsm}
In this appendix we list the useful equations for calculating the geometries in holographic WSM in section \ref{sec2}. 

The equations of motion for the action \eqref{eq:actionwsm} are 
\begin{align}
\label{eq:abeom}
\begin{split}
    R_{ab}- \frac{1}{2}g_{ab}(R+12)-T_{ab}&=0\,,\\
    \nabla_b F^{ba}+  \alpha \epsilon^{abcde} (F_{bc}F_{de} + \mathcal{F}_{bc} \mathcal{F}_{de}) - i q \left(\Phi^*(D_a\Phi) - \Phi (D_a \Phi)^*\right) &=0\,,\\
    \nabla_b  \mathcal{F}^{ba} + 2 \alpha \epsilon^{abcde} F_{bc} \mathcal{F}_{de}&=0\,,\\
    D_aD^a\Phi- m^2\Phi - \lambda \Phi^*\Phi^2&=0\,,
\end{split}
\end{align}
where
\begin{align}
     T_{ab}=&\frac{1}{2} (\mathcal F_{ac}\mathcal F_b^{~c}- \frac{1}{4} g_{ab} \mathcal F^2) + \frac{1}{2}(F_{ac}F_b^{~c}- \frac{1}{4} g_{ab} F^2) + \frac{1}{2} ((D_a\Phi)^*D_b\Phi+(D_b\Phi)^*D_a\Phi) \nn\\
  &- \frac{1}{2}g_{ab}((D_c\Phi)^*D^c\Phi+V(\Phi))\,
\end{align}
and $D_a\Phi=\partial_a\Phi - iq A_a\Phi$. 

There are three different scaling symmetries of the system
\begin{align}
  (x,y)\rightarrow a(x,y)\,, f \rightarrow a^{-2}f\,;\label{eq:scasym1}\\
  z \rightarrow az\,, h \rightarrow a^{-2}h\,, A_z \rightarrow a^{-1}A_z\,;\label{eq:scasym2}\\
  r \rightarrow ar\,, (t,x,y,z) \rightarrow a^{-1}(t,x,y,z)\,, (u,f,h) \rightarrow a^2(u,f,h)\,, A_z \rightarrow a A_z\,.\label{eq:scasym3}
\end{align}

For the ansatz \eqref{eq:ansatz}, we have equations 
\begin{align}
\label{eq:eom}
\begin{split}
  u''+\frac{h'}{2h}u'-\Big( f''+ \frac{f'h'}{2h}  \Big)\frac{u}{f}&=0\,,\\
  \frac{f''}{f}+\frac{u''}{2u} -\frac{f'^2}{4f^2} + \frac{f'u'}{fu} - \frac{6}{u} + \frac{\phi^2}{2u} \Big( m^2 + \frac{\lambda}{2}\phi^2 - \frac{q^2 A_z^2}{h} \Big) - \frac{A_z'^2}{4h} +\frac{1}{2}\phi'^2 &=0\,,\\
  \frac{1}{2}\phi'^2 +\frac{6}{u} - \frac{u'}{2u}\Big( \frac{f'}{f} + \frac{h'}{2h} \Big) - \frac{f'h'}{2fh} - \frac{f'^2}{4f^2} + \frac{A_z'^2}{4h} - \frac{\phi^2}{2u} \Big( m^2 + \frac{\lambda}{2}\phi^2 - \frac{q^2 A_z^2}{h} \Big)  &=0\,,\\
  A_z''+\Big( \frac{f'}{f} - \frac{h'}{2h}+ \frac{u'}{u} \Big)A_z' - \frac{2q^2 \phi^2}{u}A_z &=0\,,\\
  \phi'' + \Big( \frac{f'}{f}+\frac{h'}{2h}+ \frac{u'}{u} \Big)\phi' -\frac{1}{u}\, \Big( \frac{q^2 A_z^2}{h} +m^2 +\lambda \phi^2 \Big) \phi &=0\,.
\end{split}
\end{align}

Near the horizon $r=r_h$, the fields can be expanded as follows, 
\begin{align}
    \begin{split}
        u&= 4\pi T (r-r_h) + \cdots\,,\\
        f&= f_1 - f_1 A_{z2} \frac{2m^2 \phi_1^2+\lambda \phi_1^4 - 24}{6A_{z1} q^2 \phi_1^2}(r-r_h) + \cdots\,,\\
        h&= h_1 - \left( A_{z1} A_{z2} + h_1 A_{z2} \frac{2m^2 \phi_1^2+\lambda \phi_1^4 - 24}{6A_{z1} q^2 \phi_1^2} \right) (r-r_h) + \cdots\,,\\
        A_z&= A_{z1}+A_{z2}(r-r_h) + \cdots\,,\\
        \phi&= \phi_1 + A_{z2}\frac{A_{z1}^2 q^2 + h_1(m^2+\lambda \phi_1^2)}{2A_{z1} h_1 q^2 \phi_1^2}(r-r_h) + \cdots\,,
    \end{split}
\end{align}
where $T=\frac{\phi_1^2 q^2 A_{z1}}{2\pi A_{z2}}$. Note that there is a shift symmetry $r\to r+\alpha$ along the radial direction which can be used to fix $r_h$ to be any value and we choose $r_h=1$. 
There are five free parameters $T, f_1, h_1, A_{z1}, \phi_1$ and we can use the scaling symmetries (\ref{eq:scasym1}, \ref{eq:scasym2}) to fix $f_1=1, h_1=1$ respectively. Then we can shoot three 
parameters $T, A_{z1}, \phi_1$ to obtain the parameters $T, M, b$ of boundary field theory, i.e. the two 
dimensionless 
parameters $T/b, M/b$ 
according the scaling symmetry in  \eqref{eq:scasym3} (we work in unit $b=1$).    

When $r\to \infty$, the UV expansions are 
\begin{align}
\label{eq:nbwsm}
    \begin{split}
        u &= r^2- \frac{M^2}{3} + \frac{M^4(2+3\lambda)}{18}\frac{\ln r}{r^2} -\frac{M_b}{r^2} + \cdots\,,\\
        f &= r^2 - \frac{M^2}{3} + \frac{M^4(2+3\lambda)}{18}\frac{\ln r}{r^2} +\frac{f_3}{r^2} + \cdots\,,\\
        h &= r^2 - \frac{M^2}{3} + \frac{M^4(2+3\lambda)+9 b^2M^2q^2}{18} \frac{\ln r}{r^2} +\frac{h_3}{r^2} + \cdots\,,\\ 
        A_z &= b- b M^2 q^2\frac{\ln r}{r^2} + \frac{\eta}{r^2} + \cdots\,,\\ 
        \phi &= \frac{M}{r} - \frac{(3b^2 M q^2+ 2M^3+3\lambda M^3))}{6} \frac{\ln r}{r^3} + \frac{O}{r^3} + \cdots\,,
    \end{split}
\end{align}
where $h_3= \frac{1}{72} M(-72 O + 9b^2 M q^2 + M^3 (14+9\lambda)) -2f_3$. 

Note that in order to match the expansion  \eqref{eq:nbwsm} we should use the  shift symmetry of the system $r\to r+\alpha$ which could change the location of the horizon/singularity. 


\subsection{Oscillations of the scalar field in 
holographic WSM}
\label{app:osc}

In this appendix, we show that the oscillations of the scalar field obtained numerically in section \ref{sec:ist} can be analysed analytically near the singularity and the horizon. 

We focus on the case of low temperature and small $M/b$ where there are more oscillations of the scalar field. For the last equation in \eqref{eq:eom} inside the horizon, we keep the leading terms, i.e.  
\be
\label{eq:app-scalar}
\big(ufh^{1/2}\phi'\big)'=\frac{q^2A_z^2f}{h^{1/2}}\phi\,.
\ee
This equation can not be solved as a  form that $\phi$ is determined by a background phase winding. Therefore it is not similar to the Josephson effect. Nevertheless, we could use certain approximations to solve it analytically to fit the numerical data.

Near the singularity we use the approximation that $u\simeq -\frac{r_h^4}{r^2}, f\simeq r^2, h\simeq r^2, A_z\simeq A_{z1}$. Near the horizon we use the approximation that $u\simeq 4\pi T(r-r_h), f\simeq r_h^2, h\simeq r_h^2, A_z\simeq A_{z1}$.   Then the equation \eqref{eq:app-scalar} can be solved as
\be\label{eq:ssin}
\phi\simeq c_1\,J_0\Big(\frac{qA_{z1}r}{r_h^2}\Big)+
c_2\,Y_0\Big(\frac{qA_{z1}r}{r_h^2}\Big)
\ee
near the singularity and
\be\label{eq:shor}
\phi\simeq c_3\,I_0\Big(\frac{qA_{z1}}{r_h\sqrt{\pi T}}\,\sqrt{r-r_h}\Big)
\ee
near the horizon, where $J_0, Y_0$ are Bessel functions of the first kind, the second kind respectively, and $I_0$ is the modified Bessel function of the first kind. $c_i~(i=1,2,3)$ are constants which could be determined from numerical fitting. 

\begin{figure}[h!]
\begin{center}
\includegraphics[width=0.47\textwidth]{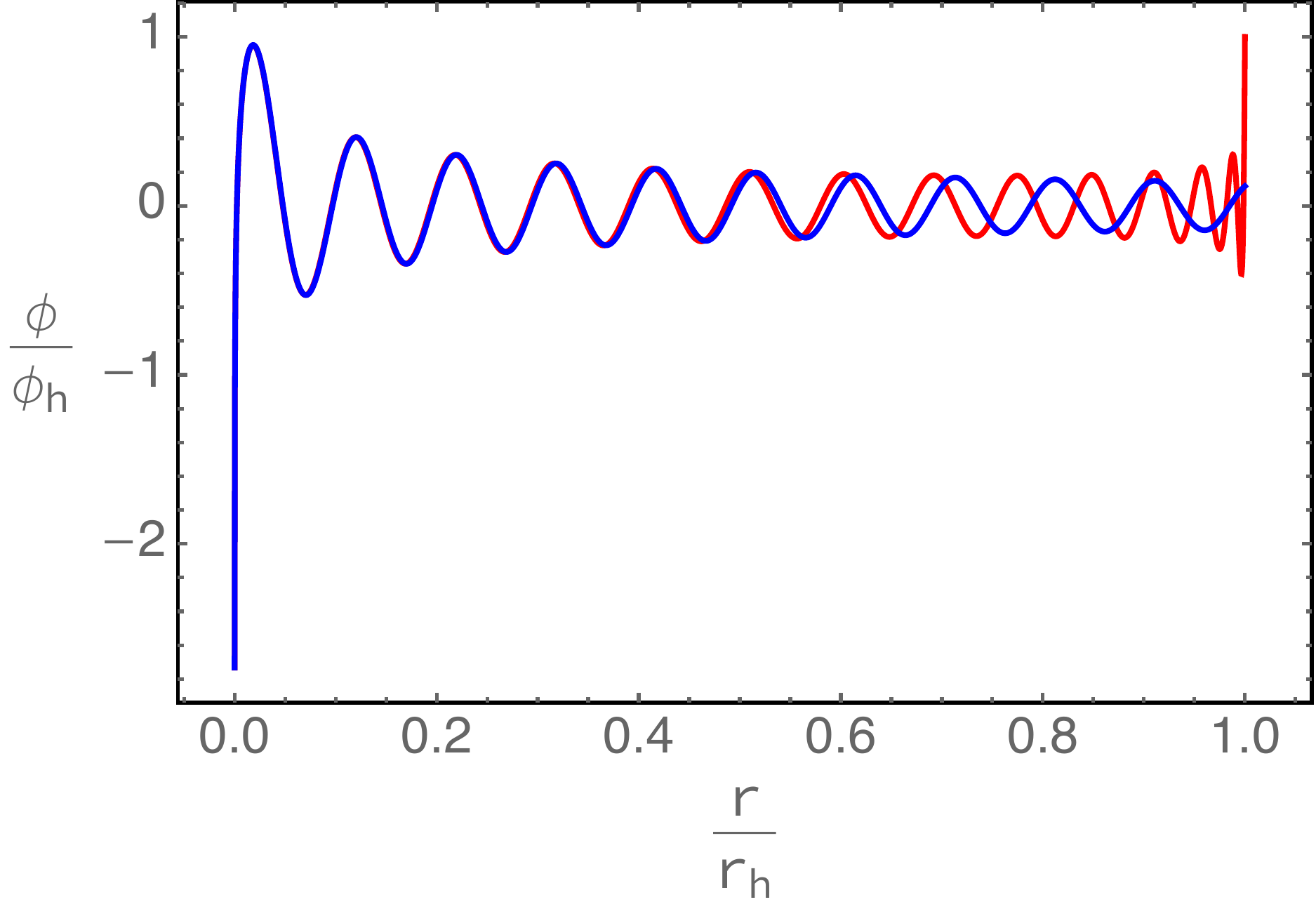}
\includegraphics[width=0.47\textwidth]{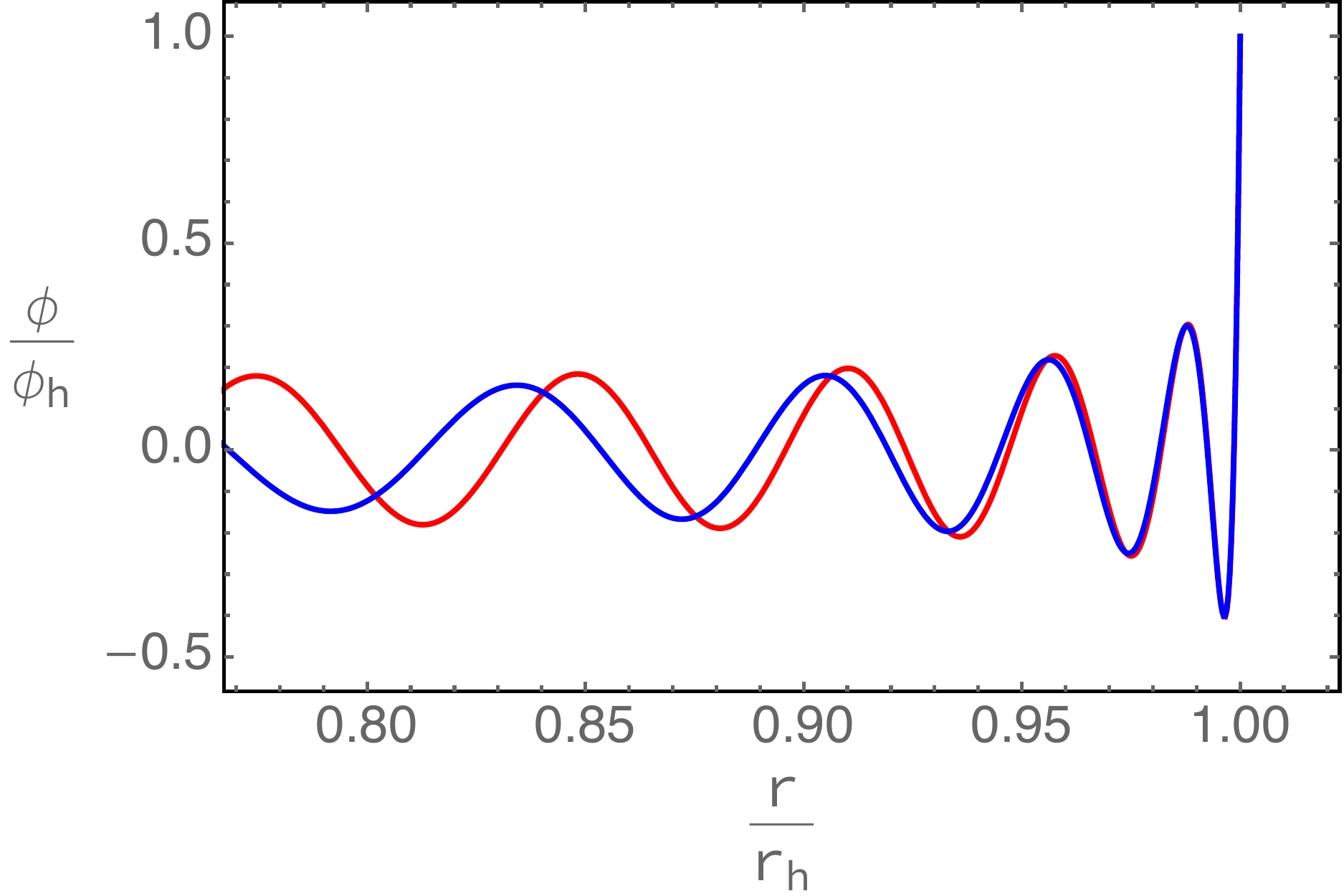}
\end{center}
\vspace{-0.4cm}
\caption{\small  Plots of the solutions for the scalar filed obtained numerically ({\em red}) and analytically ({\em blue}) near the singularity ({\em left}) and the horizon ({\em right}).}
\label{fig:sca-num-ana}
\end{figure}

We use the analytical solutions \eqref{eq:ssin} and \eqref{eq:shor} to fit the numerical solution of the scalar field at $M/b=3.03\times10^{-3}$ and $T/b=4.97\times10^{-3}$. The results are shown in Fig. \ref{fig:sca-num-ana}. We find that \eqref{eq:ssin} fits the numerical solution well in the near singularity regime, while \eqref{eq:shor} fits the numerical solution well in the near horizon regime. 

\section{Equations in holographic NLSM}
\label{app:nlsm}
In this appendix, we list the calculations for the geometries in holographic NLSM in section \ref{sec3}.  

The equations of motion for the action \eqref{eq:actionnlsm} are
\begin{align}
\begin{split}
    R_{ab}- \frac{1}{2}g_{ab}(R+12)-T_{ab}&=0\,,\\
    \nabla_b \mathcal{F}^{ba} + 2\alpha \epsilon^{abcde} F_{bc} \mathcal{F}_{de} &=0\,,\\
    \nabla_b F^{ba} + \alpha \epsilon^{abcde} (F_{bc}F_{de} + \mathcal{F}_{bc} \mathcal{F}_{de})
    -i q_1 \left(\Phi^* D^a\Phi -(D^a\Phi)^*\Phi\right) + \frac{q_2}{\eta} \epsilon^{abcde} B_{bc}B^*_{de} &=0\,,\\
    D_aD^a\Phi- \partial_{\Phi^*}V_1 - \lambda \Phi B_{ab}^*B^{ab}&=0\,,\\
    \frac{i}{3\eta} \epsilon_{abcde} H^{cde} - m_2^2B_{ab} - \lambda \Phi^*\Phi B_{ab}&=0\,,
\end{split}
\end{align}
where
\begin{align}
    T_{ab}&=\frac{1}{2} (\mathcal F_{ac}\mathcal F_b^{~c}- \frac{1}{4} g_{ab} \mathcal F^2) + \frac{1}{2}(F_{ac}F_b^{~c}- \frac{1}{4} g_{ab} F^2) + \frac{1}{2} \big((D_a\Phi)^*D_b\Phi+(D_b\Phi)^*D_a\Phi\big) \nn\\
  &+ (m_2^2 + \lambda |\Phi|^2) (B^*_{ac}B_b^{~c} + B^*_{bc}B_a^{~c}) - \frac{1}{2}g_{ab}\big((D_c\Phi)^*D^c\Phi+V_1+V_2 + \lambda |\Phi|^2 B^*_{cd}B^{cd}\big)\,.
\end{align}
With the ansatz \eqref{eq:nlansatz}, the equations are 
\begin{align}
\begin{split}
    \frac{u''}{u} - \frac{f''}{f} + \frac{h'}{2h} \left( \frac{u'}{u} - \frac{f'}{f}  \right) - \frac{4}{u}(m_2^2 + \lambda \phi^2) \left( \frac{\mathcal{B}_{tz}^2}{u h} + \frac{ \mathcal{B}_{xy}^2}{f^2} \right) &=0\,,\\
    \frac{u''}{2u} + \frac{f''}{f} - \frac{f'^2}{4f^2} + \frac{f'u'}{fu} - \frac{6}{u} + \frac{1}{u} (m_2^2 + \lambda \phi^2) \left( \frac{\mathcal{B}_{tz}^2}{u h} + \frac{ \mathcal{B}_{xy}^2}{f^2} \right) 
    ~~~~&\,\\
    + \frac{\phi^2}{2u} \left( m_1^2 + \frac{\lambda_1 \phi^2}{2} \right) +\frac{\phi'^2}{2}&=0\,,\\
    \frac{f'^2}{4f^2} +\frac{f'h'}{2fh} + \frac{u'}{2u}\left( \frac{f'}{f} + \frac{h'}{2h} \right) - \frac{6}{u}+\frac{1}{u} (m_2^2 + \lambda \phi^2) \left( -\frac{\mathcal{B}_{tz}^2}{u h} + \frac{ \mathcal{B}_{xy}^2}{f^2} \right)
    ~~~~&\,\\
    + \frac{\phi^2}{2u} \left( m_1^2 + \frac{\lambda_1 \phi^2}{2} \right)-\frac{1}{2}\phi'^2 
     &=0\,,\\
    \mathcal{B}_{tz}'- \frac{\eta \sqrt{h}}{2f}(m_2^2 + \lambda \phi^2) \mathcal{B}_{xy} &=0\,,\\
    \mathcal{B}_{xy}'- \frac{\eta f}{2\sqrt{h} u}(m_2^2 + \lambda \phi^2) \mathcal{B}_{tz} &=0\,,\\
    \phi''+ \phi' \left( \frac{u'}{u} + \frac{f'}{f} +\frac{h'}{2h} \right) - \left( m_1^2 + \lambda_1 \phi^2 - \frac{2\lambda \mathcal{B}_{tz}^2}{u h} + \frac{2\lambda \mathcal{B}_{xy}^2}{f^2} \right) \frac{\phi}{u} &=0\,.
\end{split}
\end{align}

There are three different scaling symmetries of the system 
\begin{align}
\label{eq:nscal1}
     (x,y)\rightarrow a(x,y)\,, f \rightarrow a^{-2}f\,,  \mathcal{B}_{xy}  \rightarrow a^{-2} \mathcal{B}_{xy}\,;\\
\label{eq:nscal2}
     z \rightarrow az\,, h \rightarrow a^{-2}h\,, \mathcal{B}_{tz} \rightarrow a^{-1} \mathcal{B}_{tz}\,;\\
\label{eq:nscal3}
     r \rightarrow ar\,, (t,x,y,z) \rightarrow a^{-1}(t,x,y,z)\,, (u,f,h,\mathcal{B}_{xy},\mathcal{B}_{tz}) \rightarrow a^2(u,f,h,\mathcal{B}_{xy},\mathcal{B}_{tz})\,.
\end{align}

 Near the horizon $r\to r_h$, the fields can be expanded as follows, 
\begin{align}
    \begin{split}
        u&= 4\pi T (r-r_h) + \cdots\,,\\
        f&= f_1 -  \frac{4\mathcal{B}_{xy2} \left( 8\mathcal{B}_{xy1}^2 (m_2^2+ \lambda \phi_1^2) +f_1^2 (2m_1^2\phi_1^2+\lambda_1\phi_1^4-24) \right)  }{3\mathcal{B}_{xy1} f_1 \eta^2 (m_2^2+ \lambda \phi_1^2)^2 }(r-r_h) + \cdots\,,\\
        h&= h_1 - \frac{4 h_1 \mathcal{B}_{xy2} \left( 4\mathcal{B}_{xy1}^2 (m_2^2+ \lambda \phi_1^2) - f_1^2 (2m_1^2\phi_1^2+\lambda_1\phi_1^4-24) \right)}{3\mathcal{B}_{xy1} f_1 \eta^2 (m_2^2+ \lambda \phi_1^2)^2} (r-r_h) + \cdots\,,\\
        \mathcal{B}_{xy}&= \mathcal{B}_{xy1}+\mathcal{B}_{xy2}(r-r_h) + \cdots\,,\\
        \mathcal{B}_{tz}&= \frac{\eta \sqrt{h_1} \mathcal{B}_{xy1} (m_2^2+ \lambda \phi_1^2)}{2f_1} (r-r_h) + \cdots\,,\\
        \phi&= \phi_1 + \frac{4\mathcal{B}_{xy2} \phi_1 \left( 2\lambda \mathcal{B}_{xy1}^2 +f_1^2(m_1^2+ \lambda_1 \phi_1^2) \right)}{\mathcal{B}_{xy1} f_1^2 \eta^2 (m_2^2+ \lambda \phi_1^2)^2}(r-r_h) + \cdots\,,
    \end{split}
\end{align}
where $T= \frac{\mathcal{B}_{xy1} \eta^2 (m_2^2+ \lambda \phi_1^2)^2}{16 \pi \mathcal{B}_{xy2}}$.
The strategy of the numerics is the same as  the holographic WSM. 
We first use the shift symmetry $r\to r+\alpha$ to fix $r_h=1$. Then 
we also have five free parameters $T, f_1, h_1, \mathcal{B}_{xy1}, \phi_1$ and we can use the scaling symmetries (\ref{eq:nscal1}, \ref{eq:nscal2}) to fix $f_1=1, h_1=1$ respectively. After that we have only three near horizon  
parameters $T, \mathcal{B}_{xy1}, \phi_1$, from which we obtain $T, M, b$ in the dual field theory, which are equivalently two 
dimensionless 
parameters $T/b, M/b$ according the scaling symmetry \eqref{eq:nscal3}.  

Near the boundary $r\to\infty$, we have 
\begin{align}
\label{eq:nbnlsm}
\begin{split}
    u&= r^2 - 2b^2 - \frac{M^2}{3} + \frac{8b^4+ M^4 (2+3\lambda_1)}{18} \frac{\ln r}{r^2} - \frac{M_b}{r^2} +\cdots\,,\\
    f&= r^2 - \frac{M^2}{3} + \frac{8b^4+ M^4 (2+3\lambda_1)}{18} \frac{\ln r}{r^2} +\frac{f_3}{r^2}  +\cdots\,,\\
    h&= r^2 - 2b^2 - \frac{M^2}{3} + \frac{8b^4+ M^4 (2+3\lambda_1)}{18} \frac{\ln r}{r^2} + \frac{h_3}{r^2} +\cdots\,,\\
    \mathcal{B}_{xy}&= b r + \frac{2b^3 \ln r}{r} + \frac{b_2}{r} +\cdots\,,\\
    \mathcal{B}_{tz}&= b r -  \frac{2b^3 \ln r}{r} -\frac{b\left(b^2+M^2(1+\lambda)\right)+b_2}{r} +\cdots\,,\\
    \phi &= \frac{M}{r} - \frac{M^3(2+3\lambda_1)}{6}\frac{\ln r}{r^3} + \frac{O}{r^3} +\cdots\,,
\end{split}
\end{align}
where $b_2 = \frac{1}{48b}\left(-56b^4 + 72(2f_3+h_3) - 8b^2M^2(2+3\lambda) -M^4(14+9\lambda_1) + 72MO \right)$.

Note that to match the expansion  \eqref{eq:nbnlsm} we should use the shift symmetry $r\to r+\alpha$ which could change the location of the horizon/singularity .


\end{document}